\documentclass[aps,onecolumn]{revtex4}
\usepackage{graphicx}
\usepackage{epsfig}
\usepackage{epstopdf}
\usepackage{amsfonts}
\usepackage{amssymb}
\usepackage{amsbsy}
\usepackage{amsmath}
\usepackage{mathrsfs}
\usepackage{latexsym}
\usepackage{natbib}
\usepackage{bm}
\usepackage{color}
\usepackage[a4paper, total={6.5in, 9.9in}]{geometry}

\usepackage{braket}
\usepackage{slashed}
\usepackage{pgfplots}
\numberwithin{equation}{section}

\usepackage{tikz}
\usetikzlibrary{shapes,arrows,shadows}

\begin{document}

\title{Local description of S-matrix in quantum field theory in curved 
spacetime using Riemann-normal coordinate}

\author{Susobhan Mandal}

\affiliation{Department of Physical Sciences,\\ 
Indian Institute of Science Education and Research Kolkata,\\
Mohanpur - 741 246, WB, India }

\email{sm17rs045@iiserkol.ac.in}

\author{Subhashish Banerjee}

\affiliation{Indian Institute of Technology Jodhpur, India
}
\email{subhashish@iitj.ac.in}


\begin{abstract}
The success of the S-matrix in quantum field theory in Minkowski spacetime naturally demands the extension of the construction of the S-matrix in a general curved spacetime in a covariant manner. However, it is well-known that a global description of the S-matrix may not exist in an arbitrary curved spacetime. Here, we give a local construction of S-matrix in quantum field theory in curved spacetime using Riemann-normal coordinates which mimics the methods, generally used in Minkowski spacetime. Using this construction, the scattering amplitudes and cross-sections of some scattering processes are computed in a generic curved spacetime. Further, it is also shown that these observables can be used to probe features of curved spacetime as these local observables carry curvature-dependent corrections. Moreover, the compatibility of the local construction of the S-matrix with the spacetime symmetries is also discussed in detail.
\end{abstract}

\maketitle

\section{Introduction}\label{Introduction}
S-matrix is one of the most important observables in quantum field theory (QFT), behind the discovery of nucleons to Higgs boson in different scattering phenomena in high energy particle colliders \cite{aad2012observation, cms2012observation}. The success of S-matrix in QFT raises the question regarding the description of S-matrix in QFT in curved spacetime \cite{birrell1984quantum, fulling1989aspects, wald1994quantum} where background geometry is treated as classical on which matter is quantized.

Developments of S-matrix in curved spacetime have been going on for quite some time \cite{Wald:1979wt}, though the generalization from flat Minkowski spacetime to a generic curved spacetime manifold is not straightforward. Thus, it is important to have a physical understanding of the S-matrix in curved spacetime which reproduces the known results in the flat-spacetime limit. A description of the S-matrix in curved spacetime would be deemed successful if it exhibits unitarity, analyticity, and compatibility with spacetime symmetries. A description satisfying these three conditions is not yet available \cite{stapp2004correspondence, hollowood2008causal, friedman1992unitarity}. This would provide a way to understand and place constraints on a successful theory of quantum gravity that can be probed by looking at scattering phenomena at very high energy scales where the solution of Einstein's equation $G_{\mu\nu}=T_{\mu\nu}$ would give rise to a non-trivial compact curved spacetime \cite{giddings2013gravitational}.

In \cite{Wald:1979wt}, the question of the existence of S-matrix in curved spacetime was probed, and the ``in" and ``out" creation and annihilation operators were mapped \cite{wald1975particle} through S-matrix via the Bogoliubov transformation. Here the focus was particle creation phenomena by gravitation field \cite{Wald:1979wt}. In \cite{herman1996generalized, Christensen:1978yd, Christensen:1976vb}, the point splitting method was used to regularize the divergences which arise when the separation between points in Green's function is taken to be zero. However, this approach is not useful for massless field theories as there it gives rise to divergence because of the series expansion of Green's function in $\frac{1}{\text{mass}^{2}}$. Although the point-splitting method is successful in regularizing divergence arising due to zero point-separation, the notion of particle states crucial for defining S-matrix is absent.

The S-matrix in global de-Sitter spacetime  is considered in \cite{marolf2013perturbative}. Here it was pointed out that the prescription for defining the S-matrix in flat Minkowski spacetime is not well-defined in de-Sitter space because of the IR-divergences. Hence, the Schwinger-Keldysh path integral formalism is being used with a particular contour in time integral that goes from de-Sitter horizon to past-infinity, returns to the horizon, goes all the way to future-infinity, and then comes back again to the de-Sitter horizon. In contrast, in \cite{Hofbaur:2014lfi}, it is claimed that the in/in formalism works in de-Sitter space without any divergences. Thus, there seems to be an ambiguity in defining S-matrix in de-Sitter space.

In \cite{ashtekar1975quantum}, the discussion is started by asking the question ``why, in the standard formulation of QFT, a `global' technique such as the decomposition of 
fields into positive and negative frequency parts is required for a successful description of such `local' entities as particles" which is required for describing S-matrix. The question is answered through the complex structure on the vector space of real solutions of the Klein-Gordon equation. In \cite{dohse2015complex}, it was brought out that there is no complex structure that satisfies all demands on S-matrix in AdS spacetime which are invariant under the AdS isometries, induction of a positive definite inner product, compatibility with the standard S-matrix picture, and recovery of standard structures in Minkowski spacetime under a limit of vanishing curvature. Further, it was also commented that the S-matrix construction proposal from boundary states in AdS/CFT conjecture has remained unclear due to the lack of conceptual foundation for a notion of boundary states.

On the other hand, in \cite{einhorn2003interacting}, it is suggested that since in a general curved spacetime manifold there are no asymptotic flat-regions in which the ``in" and ``out" states can be defined, it follows that there is no well-defined S-matrix in a generic curved spacetime. A similar conclusion was reached from the perspective of cosmology in \cite{bousso2005cosmology}. Here it is argued that no realistic cosmological model permits the global observations associated with an S-matrix since the observer is not outside of the system, hence, initial states can not be setup. Moreover, it is noted in \cite{banks2001m} that in a compact universe, an S-matrix description must restrict to states with a finite number of extra particles, and this restriction is necessary but not sufficient. Recently it is pointed out that the event horizon obstructs the construction of S-matrix \cite{hellerman2001string, fischler2001acceleration}. Moreover, in \cite{bousso2005cosmology}, it is also commented that there is no S-matrix in the de-Sitter universe since the observer's causal diamond misses almost all of the asymptotic regions in the global metric on which the global ``in" and ``out" states might be defined. Using the general boundary formulation, a resolution of a well-known problem of AdS is claimed in \cite{colosi2012s} where due to the lack of temporarily asymptotic free states, the usual S-matrix can not be defined. Though the unitarity of interacting field theory in curved spacetime is shown in \cite{friedman1992unitarity, giddings2013gravitational}, it is argued that unitarity of gravitational S-matrix is a more profound problem in formulating a complete theory of quantum gravity.  

Keeping in mind the above discussions and the previous works done in the literature \cite{Wald:1979wt, friedman1992unitarity, giddings2013gravitational, bousso2005cosmology, colosi2012s, spradlin2002vacuum}, we propose a technique in the present article through which a local description of S-matrix in curved spacetime can be obtained in a covariant manner using the Riemann-normal coordinate (RNC). In order to describe a global S-matrix, it is well-known \cite{wald1994quantum} that notion of asymptotic free states is required, which can be defined under the following set of conditions i) asymptotically flat spacetime or ii) non-vanishing curvature only in a compact region or iii) globally hyperbolic spacetime (which possesses a Cauchy hypersurface) which is mostly a necessary condition in the Hamiltonian formulation of General Relativity or iv) asymptotically stationary spacetime (since time-like Killing vector field gives rise to a split between positive and negative frequencies) or v) existence of Feynman propagator (propagating positive frequency in-states into the positive frequency out-states).

The second and fifth conditions mentioned above hold in our approach. We consider a local patch of spacetime manifold about a fixed point which we call the origin of that patch, in which no two geodesics intersect and where we consider scattering events to occur. This region may not be necessarily small; it depends on the background geometry such that in that region no two geodesics can intersect. The focusing of geodesics is controlled by the Ricci tensor through the Raychaudhuri equation \cite{kar2007raychaudhuri, ellis2007raychaudhuri}.  We assume that the interaction time-scale between particles is very small compared to space-like hypersurfaces that bound the compact region. In this region using the RNC, we construct states which can be treated as asymptotic states on the above-mentioned space-like hypersurfaces within the limit of our approximation. These states become exact asymptotic states in Minkowski spacetime when the curvature vanishes. Moreover, by constructing such states, we are able to reproduce Green's function in curved spacetime obtained using RNC. These states can be measured by a local observer through local measurements only \cite{book:432248}. Our construction not only enables the calculation of scattering amplitudes which captures curvature effects but is also able to show that S-matrix in curved spacetime is a local observable and scattering amplitude of a process depends on the curvature of manifolds at the origin of different chosen patches. Therefore, it varies smoothly over the whole manifold. 

In scattering amplitude measurements, measurement is done on incoming and outgoing particles having well-defined momentum states. Having globally well-defined momentum states is not possible in a generic curved spacetime because of the absence of spacetime translational Killing symmetries. However, choosing RNC patches in a generic manifold has the benefit that one can have solutions of Killing equations within the patches deformed around Poincare symmetries, shown later. Because of this, one can define well-defined momentum states within the RNC patches. However, because of deformed translational symmetries, in scattering amplitude computations, one would get terms in which net momentum of incoming and out-going particles is not conserved. Deformed symmetries can also be realized from the fact that Green's function in momentum space is explicitly not a function of the norm of 4-momentum $(k^{2})$.

The structure of the article is as follows. The first few sections are devoted to the development of the S-matrix formalism in curved spacetime. The LSZ reduction theorem in curved spacetime is reviewed in the canonical approach, and we also show the same result extending the functional integral formulation in curved spacetime. Our discussion of curved spacetime is facilitated by the use of Riemann normal coordinates, which is briefly reviewed. This is used to construct Green's function in momentum space as a series expansion, taking into account the curvature-dependent corrections following \cite{bunch1979feynman}. Using this formalism with Green's function constructed using RNC, we compute the scattering amplitudes and cross-sections in interacting field theories in curved spacetime, which is one of our main results. Moreover, we discuss how these quantities can probe geometric features of the spacetime manifolds with few examples. Further, the consistency of this formalism with S-matrix observables in terms of compatibility with spacetime symmetries is also discussed in the present article, which is our other main result. With these results, we provided a consistent and covariant way of constructing local S-matrix using RNC. 

\section{Extension of LSZ reduction theorem in curved spacetime}
The LSZ reduction formula is an important result in QFT since it states that all the S-matrix elements can be computed in principle from the correlation functions in a given field theory. In this section, we briefly review the extension of this theorem in curved spacetime. Moreover, we discuss the applicability of this theorem with an approximation in the construction of a local S-matrix in a generic curved spacetime containing no exact asymptotic quantum states. 

\subsection{Mode decomposition}
We consider the following action which describes a free scalar field theory in curved spacetime with a non-minimal coupling
\begin{equation}
S=-\int d^{4}x\sqrt{-g(x)}\Big[\frac{1}{2}g^{\mu\nu}(x)\partial_{\mu}\phi(x)\partial_{\nu}\phi(x)+\frac{1}{2}m^{2}\phi^{2}(x)+\frac{\xi}{2}\mathcal{R}(x)\phi^{2}(x)\Big],
\end{equation}
where $\xi$ denotes coupling strength between matter field $\phi$ and Ricci scalar $\mathcal{R}$. The corresponding equation of motion becomes
\begin{equation}\label{eqn.1}
(\Box-m^{2}-\xi\mathcal{R}(x))\phi(x)=0,
\end{equation}
where $\Box\phi=\frac{1}{\sqrt{-g}}\partial_{\mu}(\sqrt{-g}\partial^{\mu}\phi)=\frac{1}{\sqrt{-g}}\partial_{\mu}(\sqrt{-g}g^{\mu\nu}\partial_{\nu}\phi)$. The complete set solutions of the above equation form a basis that helps us to decompose the field 
operator as follows
\begin{equation}
\hat{\phi}(x)=\sum_{i}[\hat{a}_{i}f_{i}(x)+\hat{a}_{i}^{\dagger}f_{i}^{*}(x)],
\end{equation}
where $\{f_{i},f_{i}^{*}\}$ are the set of solutions of eqn. (\ref{eqn.1}). An inner product in functional space is defined as follows
\begin{equation}
\begin{split}
(f,g) & =-i\int_{\Sigma}d^{3}x\sqrt{-g(x)}[f^{*}(t,\vec{x})\overleftrightarrow{\partial^{0}}g(t,\vec{x})]\\
 & =-i\int_{\Sigma}d^{3}x\sqrt{-g(x)}[f^{*}(t,x)\partial^{0}g(t,\vec{x})-\partial^{0}f^{*}(t,x)g(t,\vec{x})],
\end{split}
\end{equation}
where the subscript $\Sigma$ denotes a spacelike or $t=\text{const}$ hypersurface.
If $\{f,g\}$ belongs to a set of solutions of eqn. (\ref{eqn.1}), it can be shown that this inner product is time-independent \cite{book:226340}. Therefore, using the Gram-Schmidt orthogonalization procedure, a set of solutions of eqn. (\ref{eqn.1}) can be constructed such that
\begin{equation}
(f_{i},f_{j})=\delta_{ij}, \ (f_{i}^{*},f_{j}^{*})=-\delta_{ij}, \ (f_{i},f_{j}^{*}) =(f_{i}^{*},f_{j})=0, \ \forall i,j.
\end{equation}
Using such inner products, we obtain the following relations
\begin{equation}\label{eqn.2}
\begin{split}
\hat{a}_{i}(t) & =(f_{i},\hat{\phi}(x))=-i\int d^{3}x\sqrt{-g}(f_{i}^{*}(t,\vec{x})\overleftrightarrow{\partial^{0}}\hat{\phi}(x))\\ 
\hat{a}_{i}^{\dagger}(t) & =-(f_{i}^{*},\hat{\phi}(x))=i\int d^{3}x\sqrt{-g}(f_{i}(t,\vec{x})\overleftrightarrow{\partial^{0}}\hat{\phi}(x)). 
\end{split}
\end{equation}

\subsection{Asymptotic creation, annihilation operators and LSZ reduction formula}
Using (\ref{eqn.2}), we obtain the following relation between asymptotic creation operators
\begin{equation}
\hat{a}_{i}^{\dagger}(\infty)-\hat{a}_{i}^{\dagger}(-\infty)=i\int d^{4}x\sqrt{-g}f_{i}(x)[\Box-m^{2}-\xi\mathcal{R}(x)]\hat{\phi}(x).
\end{equation}
This shows that
\begin{equation}\label{eqn.3}
\begin{split}
\hat{a}_{i}^{\dagger}(-\infty) & =\hat{a}_{i}^{\dagger}(\infty)-i\int d^{4}x\sqrt{-g}f_{i}(x)[\Box-m^{2}-\xi\mathcal{R}(x)]\hat{\phi}(x)\\
\hat{a}_{i}(\infty) & =\hat{a}_{i}(-\infty)-i\int d^{4}x\sqrt{-g}f_{i}^{*}(x)[\Box-m^{2}-\xi\mathcal{R}(x)]\hat{\phi}(x).
\end{split}
\end{equation}
In scattering amplitude calculations, these asymptotic relations will be useful, shown in the next section. These relations are similar to flat spacetime results except that we have to suitably replace the d'Alembertian operator. In scattering amplitude calculations, we are looking for quantities denoted and defined as follows
\begin{equation}
\mathcal{S}(f_{1},\ldots,f_{m}\rightarrow f_{m+1},\ldots,f_{n})=\bra{0}\mathcal{T}\hat{a}_{m+1}(\infty)
\ldots\hat{a}_{n}(\infty)\hat{a}_{1}^{\dagger}(-\infty)\ldots\hat{a}_{m}^{\dagger}(-\infty)\ket{0},
\end{equation}
where $\mathcal{T}$ is the time-ordered product operation. Making use of the relations in (\ref{eqn.3}), it can be shown that
\begin{equation}\label{LSZ theorem}
\begin{split}
\mathcal{S}(f_{1},\ldots,f_{m}\rightarrow f_{m+1},\ldots,f_{n})=\int\prod_{i=1}^{m} & \left(d^{4}z_{i}\sqrt{-g(z_{i})}\right)\prod_{j=1}^{n}\left(d^{4}y_{j}\sqrt{-g(z_{j})}\right)\\
 \times(\Box-m^{2}-\xi\mathcal{R})_{z_{1}}\ldots(\Box-m^{2}-\xi\mathcal{R})_{z_{m}} & (\Box-m^{2}-\xi\mathcal{R})_{y_{1}}\ldots(\Box-m^{2}-\xi\mathcal{R})_{y_{n}}\\
 \times \bra{0}\mathcal{T}(\hat{\phi}(z_{1})\ldots\hat{\phi}(z_{m})\hat{\phi}(y_{1})\ldots\hat{\phi}(y_{n}))\ket{0} & f_{1}(z_{1})\ldots f_{m}(z_{m})f_{m+1}^{*}(y_{1})\ldots f_{m+n}^{*}(y_{n}).
\end{split}
\end{equation}
This is the LSZ reduction formula for curved spacetime. It looks complicated from an operational point of view. There is another way of deriving the same formula using the functional integral formalism that gives a much simpler expression in terms of calculating the S-matrix elements which we discuss next.  

In order to derive the result in (\ref{LSZ theorem}), we used the relations in (\ref{eqn.3}) between the asymptotic creation and annihilation operators. Therefore, the above result would not be expected to hold in a generic curved spacetime due to the absence of asymptotic states. However, the time scale in which the interactions between particles take place during the scattering events is typically very small. Therefore, the result in (\ref{LSZ theorem}) would still hold approximately if we replace the operators $\{\hat{a}_{i}(\infty),\hat{a}_{i}(-\infty)\}$ by $\{\hat{a}_{i}(T),\hat{a}_{i}(-T)\}$ such that $2T$ is sufficiently larger than the interaction time scale. Therefore, using this approximation, computations of S-matrix elements can be done by comparing the creation and annihilation operators with some finite time difference up to some boundary term. We can impose periodic boundary conditions on $t=\pm T$ spacelike hypersurfaces such that the boundary term vanishes where $t$ is the time-like coordinate. This results in the similar relations like in (\ref{eqn.3}) except $\{\hat{a}_{i}(\infty),\hat{a}_{i}(-\infty)\}$ are replaced by $\{\hat{a}_{i}(T),\hat{a}_{i}(-T)\}$. Hence, \textit{w.r.t} that boundary condition, the incoming and outgoing states are well-defined. The spacelike hypersurfaces parametrized by $t=\pm T$ are chosen in such a way that the RNC patch is bounded by them, which is discussed later. This approximation is consistent with the construction of RNC coordinates. This motivates the construction of a local S-matrix in a generic curved spacetime using the RNC. If we consider spacetimes that have time-like Killing vector fields, the notion of in-coming and out-going states are automatically well-defined \textit{w.r.t} boundary conditions at $t\rightarrow\pm\infty$ spacelike hypersurfaces.

\section{Functional integral formulation}
In this section, we extend the functional integral formulation of field theory in curved spacetime following \cite{Nair:2005iw, seahra2000path}. Our result in this section is used later for the computation of scattering amplitudes using RNC. 

\subsection{Generating functional}
For the considered free-field theory, the generating functional can be written as
\begin{equation}
\begin{split}
\mathcal{Z}_{0}[J] & =\int\mathcal{D}\phi e^{-i\int d^{4}x\sqrt{-g}\Big[\frac{1}{2}g^{\mu\nu}\partial_{\mu}\phi\partial_{\nu}\phi+\frac{1}{2}m^{2}\phi^{2}+\frac{1}{2}\xi\mathcal{R}\phi^{2}-J\phi\Big]}\\
 & =\int\mathcal{D}\phi e^{i\int d^{4}x\sqrt{-g}\Big[\frac{1}{2}\phi(\Box-m^{2}-\xi\mathcal{R})\phi+J\phi\Big]}=\mathcal{N}' e^{\frac{i}{2}\int d^{4}x \ d^{4}y\sqrt{-g(x)}\sqrt{-g(y)}J(x)\mathcal{G}(x,y)J(y)},
\end{split}
\end{equation}
where $\mathcal{N}'=\mathcal{Z}_{0}[J=0]=\text{det}(\Box-m^{2}-\xi\mathcal{R})$. Considering $\phi^{4}$ theory as an example of interacting field theories in curved spacetime, the generating functional can be expressed as
\begin{equation}
\begin{split}
\mathcal{Z}[J] & =\mathcal{N}\int\mathcal{D}\phi e^{-i\int d^{4}x\sqrt{-g}\Big[\frac{1}{2}g^{\mu\nu}\partial_{\mu}\phi\partial_{\nu}\phi+\frac{1}{2}m^{2}\phi^{2}+\frac{1}{2}\xi\mathcal{R}\phi^{2}+\frac{\lambda}{4!}\phi^{4}-J\phi\Big]}\\
 &=\mathcal{N}e^{-i\lambda\int d^{4}x\sqrt{-g(x)}\left(\frac{1}{\sqrt{-g(x)}}\frac{\delta}{\delta J(x)}\right)^{4}}\mathcal{Z}_{0}[J].
\end{split}
\end{equation}
Here $\mathcal{N}$ is chose such a way that $\mathcal{Z}[J=0]=1$. Using the following identity
\begin{equation}
\begin{split}
F\Big[\frac{1}{\sqrt{-g}}\frac{\delta}{\delta\phi}\Big]G[\phi]e^{\int J\phi\sqrt{-g}d^{4}x}\Big|
_{\phi=0} & =F\Big[\frac{1}{\sqrt{-g}}\frac{\delta}{\delta\phi}\Big]G\Big[\frac{1}{\sqrt{-g}}
\frac{\delta}{\delta J}\Big]e^{\int J\phi\sqrt{-g}d^{4}x}\Big|_{\phi=0}=G\Big[\frac{1}{\sqrt{-g}}\frac{\delta}{\delta J}\Big]F[J],
\end{split}
\end{equation}
the generating functional can be rearranged such that
\begin{equation}
\mathcal{Z}[J]=\mathcal{N}e^{\frac{i}{2}\int d^{4}x \ d^{4}y\sqrt{-g(x)}\sqrt{-g(y)}\frac{1}{\sqrt{-g(x)}}\frac{\delta}{\delta\varphi(x)}\mathcal{G}(x,y)\frac{\delta}{\delta\varphi(y)}\frac{1}{\sqrt{-g(y)}}}e^{-i\int\sqrt{-g(z)}d^{4}z\left(\frac{\lambda\varphi^{4}}{4!}+iJ\varphi\right)}\Big|_{\varphi=0}.
\end{equation}
Using the following identity
\begin{equation}
\begin{split}
e^{\frac{i}{2}\int  d^{4}x \ d^{4}y \frac{\delta}{\delta\varphi(x)}\mathcal{G}(x,y)\frac{\delta}{\delta\varphi(y)}} & e^{\int J\varphi \sqrt{-g}d^{4}x}\\
= e^{\int J\varphi \sqrt{-g}d^{4}x} & e^{\Big[\frac{i}{2}\int_{x,y}\delta_{x}\mathcal{G}(x,y)\delta_{y}
+i\int_{x,y}\sqrt{-g(x)}J(x)\mathcal{G}(x,y)\delta_{y}+\frac{i}{2}\int_{x,y}\sqrt{-g(x)}\sqrt{-g(y)}
J(x)\mathcal{G}(x,y)J(y)\Big]},
\end{split}
\end{equation}
where
\begin{equation}
\begin{split}
\int_{x,y}\delta_{x}\mathcal{G}(x,y)\delta_{y} & \equiv\int d^{4}x \ d^{4}y\frac{\delta}{\delta\varphi(x)}\mathcal{G}(x,y)\frac{\delta}{\delta\varphi(y)}\\
\int_{x,y}\sqrt{-g(x)}J(x)\mathcal{G}(x,y)\delta_{y} & \equiv\int d^{4}x \ d^{4}y\sqrt{-g(x)}J(x)\mathcal{G}(x,y)\frac{\delta}{\delta\varphi(y)}\\
\int_{x,y}\sqrt{-g(x)}\sqrt{-g(y)}J(x)\mathcal{G}(x,y)J(y) & \equiv\int d^{4}x \ d^{4}y\sqrt{-g(x)}\sqrt{-g(y)}J(x)\mathcal{G}(x,y)J(y),
\end{split}
\end{equation}
we obtain the following expression of the generating functional
\begin{equation}\label{eqn.4}
\begin{split}
\mathcal{Z}[J] & =e^{\frac{i}{2}\int_{x,y}\sqrt{-g(x)}J(x)\mathcal{G}(x,y)\sqrt{-g(y)}J(y)}\Big[e^{i\int_{x,y}\sqrt{-g(x)}J(x)\mathcal{G}(x,y)\delta_{y}}\mathcal{F}[\varphi]\Big]_{\varphi=0}\\
\mathcal{F}[\varphi] & =\mathcal{N}e^{\frac{i}{2}\int_{x,y}\delta_{x}\mathcal{G}(x,y)\delta_{y}}e^{-i\frac{\lambda}{4!}\int_{z}\varphi^{4}\sqrt{-g}}.
\end{split}
\end{equation}

\subsection{Scattering amplitude or S-matrix}
Neglecting the first exponent of \textit{r.h.s} of eqn. (\ref{eqn.4}), we obtain the following contribution to the N-point function (the connected part of the Green's function $\mathcal{G}_{c}$)
\begin{equation}\label{eqn.5}
\begin{split}
\mathcal{G}_{c}(x_{1},\ldots,x_{N}) & =\int d^{4}z_{1}\ldots d^{4}z_{N}\mathcal{G}(x_{1},z_{1})\ldots\mathcal{G}(x_{N},z_{N})V(z_{1},\ldots,z_{N})\\
 V(z_{1},\ldots,z_{N}) & =i^{N}\frac{\delta}{\delta\varphi(z_{1})}\ldots\frac{\delta}{\delta\varphi(z_{N})}\mathcal{F}[\varphi]\Big|_{\varphi=0},
\end{split}
\end{equation}
where $\mathcal{F}[\varphi]$ is the vertex functional and $V(z_{1},\ldots,z_{N})$ is the vertex function. If $x_{1}^{0},\ldots,x_{n}^{0}\rightarrow-\infty$ and $x_{n+1}^{0},\ldots,x_{N}^{0}\rightarrow\infty$, then it can be seen that 
\begin{equation}
\{x_{1}^{0},\ldots,x_{n}^{0}\}<\{z_{1}^{0},\ldots,z_{n}^{0}\}, \ \{x_{n+1}^{0},\ldots,x_{N}^{0}\}>\{z_{n+1}^{0},\ldots,z_{N}^{0}\} .
\end{equation}
Therefore, the Green's function can be written as (from now onwards we only consider the connected Green's function, hence, we omit the subscript `c')
\begin{equation}
\begin{split}
\mathcal{G}(x_{1},\ldots,x_{N}) & =\int d^{4}z_{1}\ldots d^{4}z_{N}\sum_{\{i_{1},\ldots,i_{N}\}}
f_{i_{1}}(z_{1})f_{i_{1}}^{*}(x_{1})\ldots f_{i_{n}}(z_{n})f_{i_{n}}^{*}(x_{n})\\
 \times & f_{i_{n+1}}^{*}(z_{n+1})f_{i_{n+1}}(x_{n+1})\ldots f_{i_{N}}^{*}(z_{N})f_{i_{N}}(x_{N})
V(z_{1},\ldots,z_{N}).
\end{split}
\end{equation}
On the other hand, from the eqn. (\ref{eqn.5}), we obtain the following relation
\begin{equation}
V(z_{1},\ldots,z_{N})=\prod_{j=1}^{N}\sqrt{-g(z_{j})}(\Box-m^{2}-\xi\mathcal{R})_{z_{j}}\mathcal{G}(z_{1},\ldots,z_{N}).
\end{equation}
Hence, according to the definition, the S-matrix elements can be expressed as 
\begin{equation}\label{extra2}
\begin{split}
\mathcal{S}(f_{1},\ldots,f_{n} & \rightarrow f_{n+1},\ldots,f_{N})=\bra{f_{1},\ldots,f_{n}}\hat{S}\ket{f_{n+1},\ldots,f_{N}}\\
=\prod_{j=1}^{n}\int_{z_{j}}f_{j}(z_{j}) & \sqrt{-g(z_{j})}(\Box-m^{2}-\xi\mathcal{R})_{z_{j}}\prod_{i=n+1}^{N}\int_{z_{i}}f_{i}^{*}(z_{i})\sqrt{-g(z_{i})}(\Box-m^{2}-\xi\mathcal{R})_{z_{i}}\mathcal{G}(z_{1},\ldots,z_{N}),
\end{split}
\end{equation}
where $\hat{S}$ is S-matrix operator. The above equation brings out the LSZ reduction formula, derived from the functional integral formulation. Using the following relation
\begin{equation}
\begin{split}
\mathcal{Z}[(\Box-m^{2}-\xi\mathcal{R})\tilde{\varphi}] & =e^{\frac{i}{2}\int_{x,y}
\sqrt{-g(x)}(\Box-m^{2}-\xi\mathcal{R})_{x}\tilde{\varphi}(x)\mathcal{G}(x,y)(\Box-m^{2}-\xi\mathcal{R})_{y}\tilde{\varphi}(y)\sqrt{-g(y)}}\\
\times & \Big[e^{i\int_{x,y}\sqrt{-g(x)}(\Box-m^{2}-\xi\mathcal{R})_{x}\tilde{\varphi}(x)\mathcal{G}(x,y)\frac{\delta}{\delta\varphi(y)}}\mathcal{F}[\varphi]\Big]_{\varphi=0},
\end{split}
\end{equation}
and by doing integration by parts twice, we obtain the following relation
\begin{equation}
\mathcal{F}[\tilde{\varphi}]=e^{\frac{i}{2}\int_{x}\sqrt{-g(x)}\tilde{\varphi}(x)(\Box-m^{2}-\xi\mathcal{R})_{x}\tilde{\varphi}(x)}\mathcal{Z}[-i(\Box-m^{2}-\xi\mathcal{R})\tilde{\varphi}].
\end{equation}

\section{Riemann-normal coordinates}
From the point splitting method, it is well-known that on an underlying curved manifold, explicit calculations for a perturbative expansion of some physical observable requires a lot more effort than in flat spacetime. Hence, a formulation that allows the use of flat-spacetime techniques on curved spacetime would be very welcome. A procedure that allows one to decouple gravitation interaction from the quadratic kinetic part of the action was shown in \cite{bunch1979feynman} using Riemann-normal coordinates. This approach is different from the heat kernel method that is used in zeta-function regularization and point-splitting methods. In this section, we briefly review the Riemann-normal coordinates as a preliminary material for later studies. Moreover, we also briefly reviewed the construction of Green's function using these coordinates, following \cite{bunch1979feynman}. However, at the end of section \ref{Green's function in curved spacetime}, we provided briefly the canonical way to obtain the same expression of Green's function in curved spacetime using creation and annihilation operators and local description of particle states given in section \ref{state description}.

\subsection{Riemann-normal coordinate expansion}
Riemann-normal coordinates are the closest analogue of flat spacetime coordinates in curved spacetime. This coordinate system is defined in such a way that the geodesics emanating from one point to another point of the manifold are mapped to a straight line or in other words, the vector lies on the tangent space. As mentioned earlier this coordinate does not cover the whole manifold, rather it covers a patch in the neighborhood of a given point, which will be taken as the origin of this coordinate system. This coordinate frame is well-defined as long as the geodesics do not intersect in that patch.

The main idea behind defining RNC (Riemann-normal coordinate) expansion is to use geodesics through the origin to define geodesics at neighboring points to recreate locally a smooth patch. In this coordinate, a generic point $x$ is defined by the components of the tangent vector, evaluated at the origin of that coordinate system, to the geodesic which links $x$ and $x'$, where we choose $x'$ to be the origin of this coordinate frame. Denoting the components of the tangent vector by $z^{\mu}$ and $s$ the affine parameter of the geodesic measured between $x$ and $x'$, Riemann-normal coordinates are defined as follows
\begin{equation}
x^{\mu}=sz^{\mu}, \ z^{\mu}=-\nabla_{x}^{\mu}\sigma(x,x'),
\end{equation}  
where $\sigma(x,x')$ is the Synge's world function \cite{book:432248, cardin2004global, teyssandier2006relativistic, poisson2011motion}. Synge's world function satisfies the following identities
\begin{equation}
\begin{split}
\sigma(x,x') & =\frac{1}{2}\sigma^{;\mu}(x,x')\sigma_{;\mu}(x,x')=\frac{1}{2}\sigma^{\mu}(x,x')\sigma_{\mu}(x,x')\\
 & =\frac{1}{2}\sigma^{\mu'}(x,x')\sigma_{\mu'}(x,x')=\sigma(x',x),
\end{split}
\end{equation}
where $\mu'$ denotes derivative \textit{w.r.t} $x'^{\mu}$ and $;$ represents covariant derivative. A transformation from characterizing a geodesic by $\{x^{\mu},x'^{\mu}\}$ to $\{x^{\mu},\sigma_{\nu'}(x,x')\}$ is given 
by the Jacobian
\begin{equation}
\frac{\partial(\sigma_{\nu'},x^{\rho'})}{\partial(x_{\mu},x'^{\sigma})}=\text{det}
\begin{bmatrix}
\frac{\partial\sigma_{\nu'}}{\partial x_{\mu}} & \frac{\partial\sigma_{\nu'}}{\partial x'^{\tau}}\\
\frac{\partial x'^{\rho}}{\partial x_{\mu}} & \frac{\partial x'^{\rho}}{\partial x'^{\tau}}
\end{bmatrix}=\text{det}\frac{\partial(\sigma_{\nu'})}{\partial(x_{\mu})}=\text{det}(\sigma_{,\mu\nu'}).
\end{equation}
We define the following quantity
\begin{equation}
D_{\mu\nu'}=-\sigma_{\mu\nu'},
\end{equation}
in terms of which the van Vleck-Morette determinant defined as
\begin{equation}
D(x,x')=\text{det}(D_{\mu\nu'}),
\end{equation}
whose divergence gives us the information about a caustic surface on which two geodesics can intersect. Another geometrical quantity which would be required is given by
\begin{equation}
\Delta(x,x')\equiv g^{-\frac{1}{2}}(x)D(x,x')g^{-\frac{1}{2}}(x').
\end{equation}

\subsection{Expansion of metric in RNC}
Let us consider a generic expansion of the metric tensor in RNC upto sixth order
\begin{equation}\label{eqn.6}
\begin{split}
g_{\mu\nu}(x=x'+z) & =\eta_{\mu\nu}+A_{\mu\alpha\nu\beta}z^{\alpha}z^{\beta}+B_{\mu\alpha\nu\beta\gamma}z^{\alpha}z^{\beta}z^{\gamma}+C_{\mu\alpha\nu\beta\gamma\delta}z^{\alpha}z^{\beta}z^{\gamma}z^{\delta}\\
 & +D_{\mu\alpha\nu\beta\gamma\delta\eta}z^{\alpha}z^{\beta}z^{\gamma}z^{\eta}+E_{\mu\alpha\nu\beta\gamma\delta\eta\theta}z^{\alpha}z^{\beta}z^{\gamma}z^{\delta}z^{\eta}z^{\theta}+\mathcal{O}(z^{7}).
\end{split}
\end{equation}
Similarly for the inverse metric $g^{\mu\nu}$
\begin{equation}
\begin{split}
g^{\mu\nu}(x=x'+z) & =\eta^{\mu\nu}+A_{ \ \alpha \ \beta}^{'\mu \ \nu}z^{\alpha}z^{\beta}+B_{ \ \alpha \ \beta\gamma}^{'\mu \ \nu}z^{\alpha}z^{\beta}z^{\gamma}+C_{ \ \alpha \ \beta\gamma\delta}^{'\mu \ \nu}z^{\alpha}z^{\beta}z^{\gamma}z^{\delta}\\
 & +D_{ \ \alpha \ \beta\gamma\delta\eta}^{'\mu \ \nu}z^{\alpha}z^{\beta}z^{\gamma}z^{\delta}z^{\eta}+E_{ \ \alpha \ \beta\gamma\delta\eta\theta}^{'\mu \ \nu}z^{\alpha}z^{\beta}z^{\gamma}z^{\delta}z^{\eta}z^{\theta}+\mathcal{O}(z^{7}). 
\end{split}
\end{equation}
Here all the coefficients are evaluated at coordinate $x'$. A relation between above set of coefficients can be found from the following condition
\begin{equation}
g^{\mu\rho}(x)g_{\rho\nu}(x)=\delta_{ \ \nu}^{\mu}.
\end{equation} 
The coefficients of the expansion in eqn. (\ref{eqn.6}) can be obtained as follows.
Let us take a tensorial quantity $T_{\mu\nu}(x)$ and expanded around the origin $z=0$ as
\begin{equation}
T_{\mu\nu}(x)=\sum_{n=0}^{\infty}\frac{1}{n!}\left(\frac{\partial}{\partial z^{\alpha_{n}}}\ldots\frac{\partial}{\partial z^{\alpha_{1}}}T_{\mu\nu}\right)\Big|_{z=0}z^{\alpha_{1}}\ldots z^{\alpha_{n}},
\end{equation}
where the Taylor coefficients are also tensors belonging to tangent or cotangent space (depending on the indices) at the origin since the Riemann-normal coordinates are themselves vectors belonging to the tangent space at the origin. Using the fact that the geodesic equation is satisfied by Riemann-normal coordinates, it can be shown that in these coordinates, the Christoffel symbol at origin is zero but not necessarily its derivatives. Hence, we obtain the following relations
\begin{equation}
\begin{split}
\partial_{\alpha}T_{\mu\nu}|_{z=0} & =\nabla_{\alpha}T_{\mu\nu}|_{z=0}\\
\partial_{\alpha}\partial_{\beta}T_{\mu\nu}|_{z=0} & =\nabla_{\alpha}\nabla_{\beta}T_{\mu\nu}|_{z=0}+\partial_{\alpha}\Gamma_{ \ \beta\mu}^{\sigma}|_{z=0}T_{\sigma\nu}+\partial_{\alpha}\Gamma_{ \ \beta\nu}^{\sigma}|_{z=0}T_{\mu\sigma},
\end{split}
\end{equation}
and so on. Further, it can also be shown that
\begin{equation}\label{eqn.7}
R_{ \ \alpha\beta\gamma}^{\mu}(0)=\partial_{\beta}\Gamma_{ \ \alpha\gamma}^{\mu}(0)-\partial_{\gamma}\Gamma_{ \ \alpha\beta}^{\mu}(0).
\end{equation}
Using the fact that the symmetrized version of derivatives of Christoffel symbol at any order is zero \cite{brewin1997riemann}, we obtain the following relation
\begin{equation}
\begin{split}
R_{ \ \alpha\nu\beta}^{\mu}(0)+R_{ \ \beta\nu\alpha}^{\mu}(0) & =\partial_{\nu}\Gamma_{ \ \alpha\beta}^{\mu}(0)-\partial_{\beta}\Gamma_{ \ \alpha\nu}^{\mu}(0)+\partial_{\nu}\Gamma_{ \ \beta\alpha}^{\mu}(0)-\partial_{\alpha}\Gamma_{ \ \beta\nu}^{\mu}(0)\\
\implies\partial_{\nu}\Gamma_{ \ \alpha\beta}^{\mu}(0) & =\frac{1}{3}(R_{ \ \alpha\nu\beta}^{\mu}(0)+R_{ \ \beta\nu\alpha}^{\mu}(0)).
\end{split}
\end{equation} 
In place of $T_{\mu\nu}$ if we consider the metric $g_{\mu\nu}$, we obtain
\begin{equation}
g_{\mu\nu}(z)=\eta_{\mu\nu}-\frac{1}{3}R_{\mu\alpha\nu\beta}(0)z^{\alpha}z^{\beta}+\mathcal{O}(z^{3}).
\end{equation}
Similarly, we can obtain the higher order terms by considering the high order derivatives like in eqn (\ref{eqn.7}). As a result, it can be shown that
\begin{equation}\label{extra1}
\begin{split}
g_{\mu\nu}(z) & =\eta_{\mu\nu}-\frac{1}{3}R_{\mu\alpha\nu\beta}z^{\alpha}z^{\beta}-\frac{1}{6}R_{\mu\alpha\nu\beta;\gamma}z^{\alpha}z^{\beta}z^{\gamma}-\left(\frac{1}{20}R_{\mu\alpha\nu\beta;\gamma\delta}-\frac{2}{45}R_{ \ \alpha\beta\mu}^{\lambda}
R_{\lambda\gamma\delta\nu}\right)z^{\alpha}z^{\beta}z^{\gamma}z^{\delta}\\
 & -\left(\frac{1}{90}R_{\mu\alpha\nu\beta;\gamma\delta\eta}-\frac{2}{45}R_{ \ \alpha\beta\mu}^{\lambda}R_{\lambda\gamma\delta\nu;\eta}\right)z^{\alpha}z^{\beta}z^{\gamma}z^{\delta}z^{\eta}-\Bigg[\frac{1}{504}R_{\mu\alpha\nu\beta;\gamma\delta\eta\theta}-\frac{17}{1260}R_{ \ \alpha\beta\mu}^{\lambda}R_{\lambda\gamma\delta\nu;\eta\theta}\\
 & -\frac{11}{1008}R_{ \ \alpha\beta\mu;\eta}^{\lambda}R_{\lambda\gamma\delta\nu;\theta}-\frac{1}{315}R_{\mu\alpha\beta}^{ \ \ \ \lambda}R_{\lambda\gamma\delta}^{ \ \ \ \kappa}R_{\kappa\eta\theta\nu}\Bigg]z^{\alpha}z^{\beta}z^{\gamma}z^{\delta}z^{\eta}z^{\theta}+\mathcal{O}(z^{7})\\
g^{\mu\nu}(z) & =\eta^{\mu\nu}+\frac{1}{3}R_{ \ \alpha \ \beta}^{\mu \ \nu}z^{\alpha}z^{\beta}+\frac{1}{6}R_{ \ \alpha \ \beta;\gamma}^{\mu \ \nu}z^{\alpha}z^{\beta}z^{\gamma}+\left(\frac{1}{20}R_{ \ \alpha \ \beta;\gamma\delta}^{\mu \ \nu}+\frac{1}{15}R_{ \ \alpha\beta}^{\lambda \ \ \mu}R_{\lambda\gamma\delta}^{ \ \ \ \nu}\right)z^{\alpha}z^{\beta}z^{\gamma}z^{\delta}\\
 & +\left(\frac{1}{90}R_{ \ \alpha \ \beta;\gamma\delta\eta}^{\mu \ \nu}+\frac{1}{15}R_{ \ \alpha\beta}^{\lambda \ \ \mu}R_{\lambda\gamma\delta \ ;\eta}^{ \ \ \ \nu}\right)z^{\alpha}z^{\beta}z^{\gamma}z^{\delta}z^{\eta}+\Bigg[\frac{1}{504}R_{ \ \alpha \ \beta;\gamma\delta\eta\theta}^{\mu \ \nu}+\frac{5}{252}R_{ \ \alpha\beta}^{\lambda \ \ \mu}R_{\lambda\gamma\delta \ ;\eta\theta}^{ \ \ \ \nu}\\
 & +\frac{17}{1008}R_{ \ \alpha\beta \ ;\eta}^{\lambda \ \ \mu}R_{\lambda\gamma\delta \ ;\theta}^{\nu}-\frac{2}{189}R_{ \ \alpha\beta}^{\mu \ \ \lambda}R_{\lambda\gamma\delta}^{ \ \ \ \kappa}R_{\kappa\eta\theta}^{ \ \ \ \nu}\Bigg]z^{\alpha}z^{\beta}z^{\gamma}z^{\delta}z^{\eta}z^{\theta}+\mathcal{O}(z^{7}),\\
\end{split}
\end{equation}

\begin{align*} 
-g(z) & =1-\frac{1}{3}R_{\alpha\beta}z^{\alpha}z^{\beta}-\frac{1}{6}R_{\alpha\beta;\gamma}z^{\alpha}z^{\beta}z^{\gamma}+\left(\frac{1}{18}R_{\alpha\beta}R_{\gamma\delta}-\frac{1}{20}R_{\alpha\beta;\gamma\delta}-\frac{1}{90}R_{ \ \alpha\beta}^{\lambda \ \ \kappa}R_{\lambda\gamma\delta\kappa}\right)z^{\alpha}z^{\beta}z^{\gamma}z^{\delta}\\ 
 & +\left(\frac{1}{18}R_{\alpha\beta}R_{\gamma\delta;\eta}-\frac{1}{90}R_{\alpha\beta;\gamma\delta\eta}-\frac{1}{90}R_{ \ \alpha\beta}^{\lambda \ \ \kappa}R_{\lambda\gamma\delta\kappa;\eta}\right)z^{\alpha}z^{\beta}z^{\gamma}z^{\delta}z^{\eta}+\Bigg[\frac{1}{72}R_{\alpha\beta;\eta}R_{\gamma\delta;\theta}\\
 & -\frac{1}{504}R_{\alpha\beta;\gamma\delta\eta\theta}-\frac{1}{315}R_{ \ \alpha\beta}^{\lambda \ \ \kappa}R_{\lambda\gamma\delta\kappa;\eta\theta}-\frac{1}{336}R_{ \ \alpha\beta \ \ ;\eta}^{\lambda \ \ \kappa}R_{\lambda\gamma\delta\kappa;\theta}+\frac{1}{60}R_{\alpha\beta}R_{\gamma\delta;\eta\theta}\\
 & +\frac{2}{2835}R_{\rho\alpha\beta}^{ \ \ \ \lambda}R_{\lambda\gamma\delta}^{ \ \ \ \kappa}R_{\kappa\eta\theta}^{ \ \ \ \rho}-\frac{1}{162}R_{\alpha\beta}R_{\gamma\delta}R_{\eta\theta}-\frac{1}{30}R_{\alpha\beta}R_{ \ \gamma\delta}^{\lambda \ \ \kappa}R_{\lambda\eta\theta\kappa}\Bigg]z^{\alpha}z^{\beta}z^{\gamma}z^{\delta}z^{\eta}z^{\theta}+\mathcal{O}(z^{7}).
\end{align*}

\subsection{d'Alembertian operator in curved spacetime using RNC}
Green's function $\mathcal{G}(x,y)$ in a curved spacetime is defined as the solution of the following equation 
\begin{equation}\label{eqn.8}
(-g(x))^{\frac{1}{4}}(-\Box_{x}+m^{2}+\xi\mathcal{R}(x))\mathcal{G}(x,y)(-g(y))^{\frac{1}{4}}=-\delta^{(4)}(x-y).
\end{equation}
Let us define the symmetrized version of Green's function $\bar{\mathcal{G}}(x,y)\equiv(-g(y))^{\frac{1}{4}}\mathcal{G}(x,y)(-g(x))^{\frac{1}{4}}$. Using RNC, eqn. (\ref{eqn.8}) becomes \cite{bunch1979feynman}
\begin{equation}\label{eqn.9}
\begin{split}
\eta^{\mu\nu} & \partial_{\mu}\partial_{\nu}\bar{\mathcal{G}}-\Big[m^{2}+(\xi-\frac{1}{6})\mathcal{R}\Big]\bar{\mathcal{G}}-\frac{1}{3}R_{\alpha}^{ \ \nu}z^{\alpha}\partial_{\nu}\bar{\mathcal{G}}+\frac{1}{3}R_{ \ \alpha \ \beta}^{\mu \ \nu}z^{\alpha}z^{\beta}\partial_{\mu}\partial_{\nu}\bar{\mathcal{G}}-(\xi-\frac{1}{6})R_{;\alpha}z^{\alpha}\bar{\mathcal{G}}\\
 & +\left(\frac{1}{6}R_{\alpha\beta;}^{ \ \ \nu}-\frac{1}{3}R_{\alpha \ \ ;\beta}^{ \ \nu}\right)z^{\alpha}z^{\beta}\partial_{\nu}\bar{\mathcal{G}}+\frac{1}{6}R_{ \ \alpha \ \beta;\gamma}^{\mu \ \nu}z^{\alpha}z^{\beta}z^{\gamma}\partial_{\mu}\partial_{\nu}\bar{\mathcal{G}}-\frac{1}{2}(\xi-\frac{1}{6})R_{;\alpha\beta}z^{\alpha}z^{\beta}\bar{\mathcal{G}}\\
 & +\left(\frac{1}{40}\Box R_{\alpha\beta}-\frac{1}{120}R_{;\alpha\beta}-\frac{1}{130}R_{\alpha}^{ \ \mu}R_{\beta\mu}-\frac{1}{60}R^{\lambda\kappa}R_{\lambda\alpha\beta\kappa}+\frac{1}{60}R_{ \ \alpha}^{\lambda \ \mu\kappa}R_{\lambda\beta\mu\kappa}\right)z^{\alpha}z^{\beta}\bar{\mathcal{G}}\\
 & +\left(\frac{1}{10}R_{\alpha\beta; \ \gamma}^{ \ \ \nu}-\frac{3}{20}R_{\alpha \ ;\beta\gamma}^{ \ \nu}+\frac{1}{60}R_{ \ \alpha}^{\lambda}R_{\lambda\beta\gamma}^{ \ \ \ \nu}-\frac{1}{15}R_{ \ \alpha\beta}^{\lambda \ \ \kappa}R_{\lambda\gamma \ \kappa}^{ \ \ \nu}\right)z^{\alpha}z^{\beta}z^{\gamma}\partial_{\nu}\bar{\mathcal{G}}\\
 & +\left(\frac{1}{20}R_{ \ \alpha \ \beta;\gamma\delta}^{\mu \ \nu}+\frac{1}{15}R_{ \ \alpha\beta}^{\lambda \ \ \mu}R_{\lambda\gamma\delta}^{ \ \ \ \nu}\right)z^{\alpha}z^{\beta}z^{\gamma}z^{\delta}\partial_{\mu}\partial_{\nu}\bar{\mathcal{G}}=-\delta^{(4)}(x-y).
\end{split}
\end{equation}
From now onwards, we only deal with the symmetrized Green's function. In momentum space, we can express the Green's function as a series expansion in the following manner
\begin{equation}
\bar{\mathcal{G}}(z)=\int\frac{d^{4}k}{(2\pi)^{4}}e^{ik_{\mu}z^{\mu}}\bar{\mathcal{G}}(k).
\end{equation}
In momentum space eqn. (\ref{eqn.9}) can be solved by taking into account derivatives of the metric in ascending order as
\begin{equation}\label{extra0}
\bar{\mathcal{G}}(k)=\bar{\mathcal{G}}_{0}(k)+\bar{\mathcal{G}}_{1}(k)+\bar{\mathcal{G}}_{2}(k)+\ldots,
\end{equation}
where $\bar{\mathcal{G}}_{i}(k)$ contains $i^{\text{th}}$ derivative of metric at $z=0$ and is of the order of $k^{-(2+i)}$ such that dimension-wise all the terms match.

\subsection{Green's function in momentum space using RNC}\label{Green's function in curved spacetime}
One can immediately verify that at lowest order in which no derivative of metric arises, the only terms remaining in eqn. (\ref{eqn.9}) are 
\begin{equation}
\eta^{\mu\nu}\partial_{\mu}\partial_{\nu}\bar{\mathcal{G}}_{0}(x,y)-m^{2}\bar{\mathcal{G}}_{0}(x,y)=-\delta^{(4)}(x-y),
\end{equation}
hence, in Fourier space, we obtain
\begin{equation}
\bar{\mathcal{G}}_{0}(k)=\frac{1}{k^{2}+m^{2}}.
\end{equation}
Since there is no first order derivative of metric at $z=0$ in eqn. (\ref{eqn.9}), we obtain
\begin{equation}
\bar{\mathcal{G}}_{1}(k)=0.
\end{equation}
At second order, we obtain the following relation
\begin{equation}
\eta^{\mu\nu}\partial_{\mu}\partial_{\nu}\bar{\mathcal{G}}_{2}(x,y)-m^{2}\bar{\mathcal{G}}_{2}(x,y)-(\xi-\frac{1}{6})\mathcal{R}\bar{\mathcal{G}}_{0}(x,y)-\frac{1}{3}R_{ \ \alpha}^{\nu}z^{\alpha}\partial_{\nu}\bar{\mathcal{G}}_{0}(x,y)+\frac{1}{3}R_{ \ \alpha \ \beta}^{\mu \ \nu}z^{\alpha}z^{\beta}\partial_{\mu}\partial_{\nu}\bar{\mathcal{G}}_{0}(x,y)=0.
\end{equation}
It is important to recall that the term $\bar{\mathcal{G}}_{0}(x,y)$ depends manifestly only on the scalar $z^{2}=\eta_{\mu\nu}z^{\mu}z^{\nu}$. For this kind of function, we have the following simple identity
\begin{equation}\label{eqn.10}
-\frac{1}{3}R_{ \ \alpha}^{\nu}z^{\alpha}\partial_{\nu}\bar{\mathcal{G}}_{0}(x,y)+\frac{1}{3}R_{ \ \alpha \ \beta}^{\mu \ \nu}z^{\alpha}z^{\beta}\partial_{\mu}\partial_{\nu}\bar{\mathcal{G}}_{0}(x,y)=0,
\end{equation}
using which we obtain
\begin{equation}
\eta^{\mu\nu}\partial_{\mu}\partial_{\nu}\bar{\mathcal{G}}_{2}(x,y)-m^{2}\bar{\mathcal{G}}_{2}(x,y)=(\xi-\frac{1}{6})\mathcal{R}\bar{\mathcal{G}}_{0}(x,y)\implies\bar{\mathcal{G}}_{2}(k)=-\frac{(\xi-\frac{1}{6})\mathcal{R}}{(k^{2}+m^{2})^{2}}.
\end{equation}
Similarly, because of the $z^{2}$ dependence of $\bar{\mathcal{G}}_{2}$, we obtain the following identities
\begin{equation}\label{eqn.11}
\begin{split}
\left(\frac{1}{6}R_{\alpha\beta;}^{ \ \ \ \nu}-\frac{1}{3}R_{\alpha \ ;\beta}^{ \ \nu}\right) & z^{\alpha}z^{\beta}\partial_{\nu}\bar{\mathcal{G}}_{0}+\frac{1}{6}R_{ \ \alpha \ \beta;\gamma}^{\mu \ \nu}z^{\alpha}z^{\beta}z^{\gamma}\partial_{\mu}\partial_{\nu}\bar{\mathcal{G}}_{0}=0\\
\Big[\frac{1}{10}R_{\alpha\beta;\gamma}^{ \ \ \ \nu}-\frac{3}{20}R_{\alpha \ ;\beta\gamma}^{ \ \nu} & +\frac{1}{60}R_{ \ \alpha}^{\lambda}R_{\lambda\beta\gamma}^{ \ \ \ \nu} -\frac{1}{15}R_{ \ \alpha\beta}^{\lambda \ \ \kappa}R_{\lambda\gamma \ \kappa}^{ \ \ \nu}\Big]z^{\alpha}z^{\beta}z^{\gamma}\partial_{\nu}\bar{\mathcal{G}}_{0}\\
 & +\left(\frac{1}{20}R_{ \ \alpha \ \beta;\gamma\delta}^{\mu \ \nu}+\frac{1}{15}R_{ \ \alpha\beta}^{\lambda \ \ \mu}R_{\lambda\gamma\delta}^{ \ \ \ \nu}\right)z^{\alpha}z^{\beta}z^{\gamma}z^{\delta}\partial_{\mu}\partial_{\nu}\bar{\mathcal{G}}_{0}=0,
\end{split}
\end{equation}
which leads to 
\begin{equation}
\eta^{\mu\nu}\partial_{\mu}\partial_{\nu}\bar{\mathcal{G}}_{3}-m^{2}\bar{\mathcal{G}}_{3}-(\xi-\frac{1}{6})R_{;\alpha}z^{\alpha}\bar{\mathcal{G}}_{0}=0.
\end{equation}
In momentum space, we obtain the following relation
\begin{equation}
(k^{2}+m^{2})\bar{\mathcal{G}}_{3}(k)-i(\xi-\frac{1}{6})R_{;\alpha}\partial^{\alpha}(k^{2}+m^{2})^{-1}=0\implies\bar{\mathcal{G}}_{3}(k)=i\frac{(\xi-\frac{1}{6})R_{;\alpha}}{k^{2}+m^{2}}\partial^{\alpha}(k^{2}+m^{2})^{-1}.
\end{equation}
At the fourth order, we are left with the following relation
\begin{equation}
\begin{split}
\eta^{\mu\nu}\partial_{\mu}\partial_{\nu}\bar{\mathcal{G}}_{4}-m^{2}\bar{\mathcal{G}}_{4}-(\xi-\frac{1}{6})\mathcal{R}\bar{\mathcal{G}}_{2} & -\frac{1}{2}(\xi-\frac{1}{6})R_{;\alpha\beta}z^{\alpha}z^{\beta}\bar{\mathcal{G}}_{0}+a_{\alpha\beta}z^{\alpha}z^{\beta}\bar{\mathcal{G}}_{0}=0\\
a_{\alpha\beta}=\frac{1}{40}\Box R_{\alpha\beta}-\frac{1}{120}R_{;\alpha\beta} & -\frac{1}{30}R_{\alpha}^{ \ \mu}R_{\beta\mu}-\frac{1}{60}R^{\lambda\kappa}R_{\lambda\alpha\beta\kappa}+\frac{1}{60}R_{ \ \alpha}^{\lambda \ \mu\kappa}R_{\lambda\beta\mu\kappa},
\end{split}
\end{equation}
that leads to the following expression in momentum space
\begin{equation}
\bar{\mathcal{G}}_{4}(k)=(\xi-\frac{1}{6})^{2}\mathcal{R}^{2}(k^{2}+m^{2})^{-3}+\frac{1}{2}(\xi-\frac{1}{6})\frac{R_{;\alpha\beta}}{k^{2}+m^{2}}\partial^{\alpha}\partial^{\beta}(k^{2}+m^{2})^{-1}-a_{\alpha
\beta}(k^{2}+m^{2})^{-1}\partial^{\alpha}\partial^{\beta}(k^{2}+m^{2})^{-1}.
\end{equation}
Similar series expansion for Green's function for fermionic and other bosonic fields can also be derived \cite{bukhbinder1983renormalization, bukhbinder1983one, bukhbinder1984local, inagaki1997dynamical, inagaki1999dynamical} which are useful in deriving effective action in curved spacetime. The identities (\ref{eqn.10}), (\ref{eqn.11}) follow from the adiabatic order expansion of the following relation which we demand
\begin{equation*}
\partial_{\mu}(g^{\mu\nu})\partial_{\nu}\bar{\mathcal{G}}(z,z')+(g^{\mu\nu}-\eta^{\mu\nu})\partial_{\mu}\partial_{\nu}\bar{\mathcal{G}}(z,z')=0,
\end{equation*}
$\frac{1}{k^{2}+m^{2}}$ expansion of Green's function brings out the existence of poles at $k^{2}=-m^{2}$. Since the poles of Green's function denote single-particle states (also can be checked in Minkowski spacetime), $k^{0}=\sqrt{\vec{k}^{2}+m^{2}}$ are justified as single-particle momentum modes in this local construction of Green's function. However, if the series expansion is resumed in a suitable manner, then the exact location of the pole might be different from $k^{2}=-m^{2}$. Moreover, the dispersion of single-particle energy could also be different. Here we restrict our discussion to the series expansion form of the Green's function.  

Although in the present article, we compute the S-matrix elements using the functional integral formalism, there must also be a canonical formalism (using creation-annihilation operators) which leads to the same results. This is briefly discussed here. This also provides the vacuum state $\ket{0}$ which is used earlier in deriving the LSZ reduction formula. We can decompose the field operator as 
\begin{equation}\label{canonical decomposition}
\hat{\phi}(x)=\int\frac{d^{3}k}{\sqrt{(2\pi)^{3}}}[\hat{a}_{\vec{k}}g_{\vec{k}}(x)+\hat{a}_{\vec{k}}^{\dagger}g_{\vec{k}}^{*}(x)],
\end{equation}
where $\{g_{\vec{k}}(x)\}$ are the momentum mode solutions of the Klein-Gordon equation locally. As a result, it is also the solution of the Klein-Gordon equation at the origin. The only non-vanishing commutation relation that the creation-annihilation operators satisfy is $[\hat{a}_{\vec{k}}, \hat{a}_{\vec{k}'}^{\dagger}]=\delta^{(3)}(\vec{k}-\vec{k}')$. We, therefore, consider local mode solutions of the form $g_{k}(x)=\frac{e^{ik.x}}{\sqrt{2\omega_{\vec{k}}}}Z_{k}$ where $Z_{k}$ is the momentum-dependent weight function. However, this weight function can be absorbed in the creation-annihilation operators, which makes mode functions normalized. As a result, the expressions in the equations (\ref{eqn.3}) and (\ref{LSZ theorem}) remain the same.  The vacuum state $\ket{0}$ is chosen such that it is annihilated by the operation of annihilation operators of all momentum modes. Using the definition of Green's function $\bra{0}\mathcal{T}(\hat{\phi}(x)\hat{\phi}(y))\ket{0}$ ($\mathcal{T}$ is the time-ordering operation), we obtain the following expression of Green's function in momentum space
\begin{equation}
\bar{\mathcal{G}}(k)=\frac{|Z_{k}|^{2}}{k^{2}+m^{2}}.
\end{equation}
Comparing the above expression with the series expansion of the Green's function obtained using the RNC, we obtain the $\frac{1}{k^{2}+m^{2}}$ series expansion of $Z_{k}$ up to an overall phase factor. In the next section, we discuss the local description of state space in the RNC patch, as we see the functions $g_{k}(x)=\frac{e^{ik.x}}{\sqrt{2\omega_{\vec{k}}}}Z_{k}$ are not the exact solutions of the Klein-Gordon equation in curved spacetime in the RNC patch.

\subsection{Local description of state space in curved spacetime using RNC}\label{state description}
In an attempt to extend all the important features of QFT in Minkowski spacetime to a general curved spacetime in a covariant manner, the notion of particle-like states must be well-defined at least under certain conditions as mentioned earlier. In the presence of curvature, state description is not well-defined globally since the Hilbert space description is attached to every point in the manifold through a map from its tangent space to vector space that depends on the quantization prescription \cite{drechsler1996quantum, graudenz1997quantum, graudenz1994space, prugovecki1996quantum}. However, the local description of the Hilbert space or particle-like states space is important since through the states and observables (which are quantum operators, also defined locally), we obtain the expectation values that are attached to the points on the manifold. Such information or observation cannot be compared between two spacetime points naturally because it depends on the choice of the path of the observer through which parallel transport procedure is performed. However, it is independent of the choice of coordinates. 

We define the single-particle momentum states through a mapping from tangent space at origin $x'$ to the Hilbert space attached to it and denote the states by $\{\ket{k}_{x'}\}$. Similarly, we define single particle position states in RNC frame through a map from spacetime manifold to tangent space at the origin through RNC coordinate defined earlier, then map those vectors to Hilbert space at the origin, denoted by $\{\ket{z}_{x'}\}$. The inner product between these states are defined as 
\begin{equation}
\braket{z|k'}=e^{ik'.z}\equiv f_{k'}(z),
\end{equation}
since in RNC locally within that patch, the metric takes the form as in eqn. (\ref{extra1}) which is locally flat. Since within the RNC patch all points are mapped through the tangent vectors at origin $x'$, we omit the subscript $x'$ from state notation.

It is quite clear that $e^{ik.z}$ is not the solution of the on-shell equation at every point within the RNC patch. However, such states can be constructed at the origin $x'$ since the Christoffel symbol at that point vanishes, which follows from the definition of the RNC patch. However, if we decompose the field operators in terms of these harmonic functions with a momentum-dependent weight function $Z_{k}$, we are able to reproduce the Green's function obtained using the RNC patch. Moreover, this weight function also depends on the curvature of spacetime. This shows that the particle states in curved spacetime form a wavepacket in its position space representation due to the non-trivial weight function $Z_{k}$ in momentum space. Even at the level of free-field theory, it is shown that the coupling between matter and spacetime geometry changes the single particle states in Minkowski spacetime represented by the Dirac-delta functions in position space. 

In local S-matrix formulation, the main idea is that before the scattering events take place at origin $x'$, momentum vectors of many-particle states can be measured locally \cite{bunch1981local}. These states belong to the Fock space constructed from the single-particle states through the direct sum of vector spaces made out of the direct product of a stack of single-particle Hilbert state spaces. Measurement of such local states can be possible since the time scale of interaction between multi-particles would be very small compared to the time-coordinate of space-like hypersurfaces that bound the small patch which can be described through RNC. The boundary of the patch can either be found out by a set of points where the van Vleck-Morette determinant \cite{visser1993van} is non-divergent or by solving Raychaudhuri equation for geodesic congruences \cite{mohajan2013space, dasgupta2012geodesic, kuniyal2015geodesic, nandan2017geodesic} in which one needs to take a collection of vector fields along geodesics and evolve them to see whether their cross-section decreases to zero or not. The  RNC description is not well-defined beyond the boundary at which it goes to zero. This computation has been done numerically for specific spacetime geometries \cite{mohajan2013space, dasgupta2012geodesic, kuniyal2015geodesic, nandan2017geodesic}.

For the algebraic or axiomatic quantum field theoretical description of local S-matrix in curved spacetime, one may see \cite{hollands2003renormalization, hollands2015quantum}. In order to define such a local description, the time-ordered operation can be restricted within the constant time hypersurfaces (space-like hypersurfaces) that effectively bound the patch \cite{hollands2002existence, hollands2001local}.

\subsection{Causality}
Mathematically, micro-causality in any field theory demands that the commutation of two field operators must vanish if they are space-like separated. This means
\begin{equation}\label{eqn.19}
[\hat{\phi}(x),\hat{\phi}(x')]=0, \ \text{if} \ (x-x')^{2}>0.
\end{equation}
This is another constraint that needs to be satisfied if one uses canonical quantization and from the functional integral point of view an equivalent statement can be obtained by demanding the following
\begin{equation}
\langle\hat{\phi}(x)\hat{\phi}(x')\hat{\mathcal{O}}_{1}(x_{1})\ldots\hat{\mathcal{O}}_{n}(x_{n})\rangle=\langle\hat{\phi}(x')\hat{\phi}(x)\hat{\mathcal{O}}_{1}(x_{1})\ldots\hat{\mathcal{O}}_{n}(x_{n})\rangle.
\end{equation}
Note that micro-causality will always be true; to explicitly show that one needs to solve the Klein-Gordon equation exactly and define the exact eigen-modes. On the other hand, our goal is not to look at the mode solutions of the Klein-Gordon equations, and it is not always possible to find analytic solutions of such complicated partial differential equations unless the underlying spacetime has special symmetries. Causality can only be exactly verified once the exact mode solutions are available.

However, the exact modes solutions of the Klein-Gordon equation can always be re-written in momentum basis through the Fourier transformation as a change of basis which should not change the micro-causality condition since eqn. (\ref{eqn.19}) is a basis independent mathematical statement. We choose the momentum basis in order to define the scattering matrix elements in terms of measurements of momentum modes which actually gives much more information, shown below.

\section{Illustration of formalism: Scattering amplitude calculations}
As an application of the developed formalism for local S-matrix in curved spacetime, the scattering amplitudes and cross-sections in interacting field theories in curved spacetime are computed in this section using the RNC. Since these interacting field theories are coupled to curved spacetime, the observables in these field theories are expected to have curvature-dependent corrections, extracted using the RNC. This is one of our main results in the present article. 

\subsection{Structure of local S-matrix}
Using the necessary information, discussed above, in eqn. (\ref{extra2}) we obtain
\begin{equation}
\begin{split}
\mathcal{S}(k_{1},\ldots,k_{n}\rightarrow k_{n+1},\ldots,k_{N}) & =\int d^{4}z_{1}\ldots d^{4}z_{N}f_{1}(z_{1})\ldots f_{n}(z_{n})f_{n+1}^{*}(z_{n+1})\ldots f_{N}^{*}(z_{N})V(z_{1},\ldots,z_{N})\\
 & =\int d^{4}z_{1}\ldots d^{4}z_{N} \ e^{i\sum_{i=1}^{n}k_{i}.z_{i}}e^{-i\sum_{j=n+1}^{N}k_{j}.z_{j}}V(z_{1},\ldots,z_{N}),
\end{split}
\end{equation}
where $V(z_{1},\ldots,z_{N})$ carries the information about the Green's function. This S-matrix is a local observable since the exponents carry information of the origin $x'$ through $z_{i}^{\mu}=\sigma^{\mu'}(x_{i},x')$ and curvature at $x'$.

We choose the matter to be conformally coupled to the Ricci scalar. This can be done by taking $\xi=\frac{1}{6}$ in the two-point or Green's function leading to the following form
\begin{equation}
\begin{split}
\bar{\mathcal{G}}(k) & =\frac{1}{k^{2}+m^{2}}-a_{\alpha\beta}(k^{2}+m^{2})^{-1}\partial^{\alpha}\partial^{\beta}(k^{2}+m^{2})^{-1}\\
a_{\alpha\beta} & =\frac{1}{40}\Box R_{\alpha\beta}-\frac{1}{120}R_{;\alpha\beta}-\frac{1}{30}R_{\alpha}^{ \ \mu}R_{\beta\mu}-\frac{1}{60}R^{\lambda\kappa}R_{\lambda\alpha\beta\kappa}+\frac{1}{60}R_{ \ \alpha}^{\lambda}R_{\lambda\beta\mu\kappa}.
\end{split}
\end{equation} 

\subsection{$3\rightarrow3$ scattering in $\phi^{4}$ theory}
In $3\rightarrow3$ scattering, we are looking to compute the following amplitude
\begin{equation}
\begin{split}
\mathcal{S}(k_{1},k_{2},k_{3}\rightarrow k_{4},k_{5},k_{6}) & =\int d^{4}z_{1}\ldots d^{4}z_{6}f_{k_{1}}(z_{1})\ldots f_{k_{3}}(z_{3})f_{k_{4}}^{*}(z_{4})\ldots f_{k_{6}}^{*}(z_{6})V(z_{1},\ldots,z_{6})\\
V(z_{1},\ldots,z_{6}) & =\frac{\delta}{\delta\varphi(z_{1})}\ldots\frac{\delta}{\delta\varphi(z_{6})}\mathcal{F}[\varphi]|_{\varphi=0}, \ \mathcal{F}[\varphi]=\mathcal{N}e^{\frac{i}{2}\int_{x,y}\delta_{x}\mathcal{G}(x,y)\delta_{y}}e^{-i\frac{\lambda}{4!}\int_{z}\sqrt{-g}\varphi^{4}}.
\end{split}
\end{equation}
Henceforth, we will not write the factor $\mathcal{N}$ as this cancels by taking $\varphi=0$ at the end. Consider only the $\varphi^{6}$ terms from $\mathcal{F}[\varphi]$ in leading order, we obtain
\begin{equation}
\begin{split}
\frac{1}{2}\int_{z_{1},z_{2}}\delta_{z_{1}}\mathcal{G}(z_{1},z_{2})\delta_{z_{2}} & \Big[-\frac{\lambda^{2}}{(4!)^{2}2!}\Big\{\int_{x}\varphi^{4}\sqrt{-g}\Big\}^{2}\Big]\\
=-\frac{\lambda^{2}}{144}\int_{z_{1},z_{2},x,y}\Big[\bar{\mathcal{G}}(z_{1},z_{2}) & \varphi^{3}(x)\delta^{(4)}(x-z_{1})\delta^{(4)}(y-z_{2})\varphi^{3}(y)(1-\frac{1}{12}R_{\alpha\beta}x^{\alpha}x^{\beta})(1-\frac{1}{12}R_{\alpha\beta}y^{\alpha}y^{\beta})\Big]\\
-\frac{\lambda^{2}}{192}\int_{z_{1},z_{2},x,y}\Big[\bar{\mathcal{G}}(z_{1},z_{2}) & \varphi^{2}(x)\delta^{(4)}(x-z_{1})\delta^{(4)}(x-z_{2})\varphi^{4}(y)(1-\frac{1}{12}R_{\alpha\beta}x^{\alpha}x^{\beta})(1-\frac{1}{12}R_{\alpha\beta}y^{\alpha}y^{\beta})\Big]\\
-\frac{\lambda^{2}}{192}\int_{z_{1},z_{2},x,y}\Big[\bar{\mathcal{G}}(z_{1},z_{2}) & \varphi^{4}(x)\delta^{(4)}(y-z_{1})\delta^{(4)}(y-z_{2})\varphi^{2}(y)(1-\frac{1}{12}R_{\alpha\beta}x^{\alpha}x^{\beta})(1-\frac{1}{12}R_{\alpha\beta}y^{\alpha}y^{\beta})\Big],
\end{split}
\end{equation}
which can be re-written as 
\begin{equation}
\begin{split}
-\frac{\lambda^{2}}{144}\int_{x,y}\Big[\bar{\mathcal{G}}(x,y) & \varphi^{3}(x)\varphi^{3}(y)(1-\frac{1}{12}R_{\alpha\beta}x^{\alpha}x^{\beta})(1-\frac{1}{12}R_{\alpha\beta}y^{\alpha}y^{\beta})\Big]\\
-\frac{\lambda^{2}}{192}\int_{x,y}\Big[\bar{\mathcal{G}}(x,y) & \varphi^{2}(x)\varphi^{4}(y)(1-\frac{1}{12}R_{\alpha\beta}x^{\alpha}x^{\beta})(1-\frac{1}{12}R_{\alpha\beta}y^{\alpha}y^{\beta})\Big]\\
-\frac{\lambda^{2}}{192}\int_{x,y}\Big[\bar{\mathcal{G}}(x,y) & \varphi^{4}(x)\varphi^{2}(y)(1-\frac{1}{12}R_{\alpha\beta}x^{\alpha}x^{\beta})(1-\frac{1}{12}R_{\alpha\beta}y^{\alpha}y^{\beta})\Big].
\end{split}
\end{equation}
Among the above three pieces, we choose the connected one since the disconnected pieces are cancelled out due to $\mathcal{N}^{-1}$.
\begin{figure}
\begin{center}
\includegraphics[height=4cm,width=4cm]{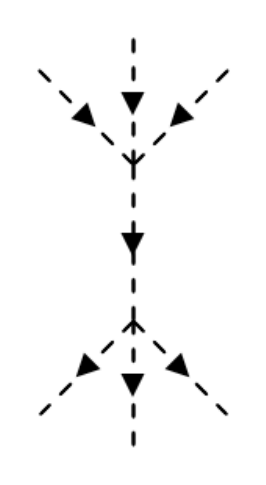}
\end{center}
\caption{Leading order connected diagram in $3\rightarrow3$ scattering in $\phi^{4}$ theory}
\end{figure}
This can be expressed as 
\begin{equation}\label{eqn.20}
\begin{split}
\int d^{4}z_{1}\ldots d^{4}z_{6}d^{4}x d^{4}y\Big[\delta^{(4)}(x-z_{1}) & \delta^{(4)}(x-z_{2})\delta^{(4)}(x-z_{3})\delta^{(4)}(y-z_{4})\delta^{(4)}(y-z_{5})\delta^{(4)}(y-z_{6})(3!)^{2}\\
+\delta^{(4)}(x-z_{1})\delta^{(4)}(x-z_{5}) & \delta^{(4)}(x-z_{3})\delta^{(4)}(y-z_{2})\delta^{(4)}(y-z_{3})\delta^{(4)}(y-z_{6})(3!)^{2}\\
+\text{other combinations}\Big]\bar{\mathcal{G}}(x,y) & (1-\frac{1}{12}R_{\alpha\beta}x^{\alpha}x^{\beta})(1-\frac{1}{12}R_{\alpha\beta}y^{\alpha}y^{\beta})\left(-\frac{\lambda^{2}}{144}\right).
\end{split}
\end{equation}
Considering a particular combination, say the first one, leads to the following contribution to the scattering amplitude expression, denoted by $\mathcal{S}^{(1)}$
\begin{equation}
\begin{split}
\mathcal{S}^{(1)}(k_{1},k_{2},k_{3}\rightarrow k_{4},k_{5},k_{6}) & =-\frac{\lambda^{2}}{4}\int d^{4}x \ d^{4}y\frac{e^{i(k_{1}+k_{2}+k_{3}).x-i(k_{4}+k_{5}+k_{6}).y}}{\prod_{i=1}^{6}\sqrt{2\omega_{k_{i}}V}}\bar{\mathcal{G}}(x,y)\\
 & \times\left(1-\frac{1}{12}R_{\alpha\beta}x^{\alpha}x^{\beta}\right)\left(1-\frac{1}{12}R_{\alpha\beta}y^{\alpha}y^{\beta}\right).
\end{split}
\end{equation}
In momentum space, the above expression becomes
\begin{equation}\label{eqn.21}
\begin{split}
\mathcal{S}^{(1)}(k_{1},k_{2},k_{3}\rightarrow k_{4},k_{5},k_{6}) & =-\frac{\lambda^{2}}{4}\int d^{4}x \ d^{4}y \ d^{4}k\frac{e^{i(k_{1}+k_{2}+k_{3}-k).x-i(k_{4}+k_{5}+k_{6}-k).y}}{\prod_{i=1}^{6}\sqrt{2\omega_{k_{i}}V}}\bar{\mathcal{G}}(k)\\
 & \times\Bigg[1-\frac{1}{12}R_{\alpha\beta}(x^{\alpha}x^{\beta}+y^{\alpha}y^{\beta})+\frac{1}{144}R_{\alpha\beta}R_{\mu\nu}x^{\mu}x^{\nu}y^{\alpha}y^{\beta}\Bigg]\\
 & =-\frac{\lambda^{2}}{4}\left(1+\frac{1}{12}R_{\alpha\beta}\partial_{k_{1}}^{\alpha}\partial_{k_{1}}^{\beta}+\frac{1}{12}R_{\alpha\beta}\partial_{k_{4}}^{\alpha}\partial_{k_{4}}^{\beta}+\frac{1}{144}R_{\mu\nu}R_{\alpha\beta}\partial_{k_{1}}^{\alpha}\partial_{k_{1}}^{\beta}\partial_{k_{4}}^{\mu}\partial_{k_{4}}^{\nu}\right)\\
 & \times\Bigg[\frac{\bar{\mathcal{G}}(k_{1}+k_{2}+k_{3})}{\prod_{i=1}^{6}\sqrt{2\omega_{k_{i}}V}}\delta^{(4)}(k_{1}+k_{2}+k_{3}-k_{4}-k_{5}-k_{6})\Bigg],
\end{split}
\end{equation}
where $V$ is the spatial volume of local patch within which the RNC description is well-defined. In a similar fashion, contribution from remaining pieces in eqn. (\ref{eqn.20}) can be calculated leading to the complete S-matrix.

We want to emphasize here that in the last equality of eqn. (\ref{eqn.21}), the last two pieces are non-conservative pieces where momentum conservation does not hold in the scattering process because of the presence of the derivative of the Dirac-delta function (see the second last equality in eqn. (\ref{eqn.21})). It would be pertinent to point out here that the coefficients of these two pieces depend on the curvature of spacetime. Hence, in the limit of flat-spacetime, these terms cease to exist. Further,  the first term in the last equality of eqn. (\ref{eqn.21}) that preserves conservation of 4-momentum also contains curvature dependent correction in the Green's function 
\begin{equation}
a_{\alpha\beta}(k^{2}+m^{2})^{-1}\partial^{\alpha}\partial^{\beta}(k^{2}+m^{2})^{-1}\Big|_{k_{1}+k_{2}+k_{3}}=\Big[-\frac{2a_{\alpha}^{ \ \alpha}}{(k^{2}+m^{2})^{3}}+\frac{8k^{\alpha}k^{\beta}a_{\alpha\beta}}{(k^{2}+m^{2})^{4}}\Big]_{k=k_{1}+k_{2}+k_{3}}.
\end{equation}
For massless scalar field theory the above term becomes
\begin{equation}
\Big[-\frac{2a_{\alpha}^{ \ \alpha}}{(k^{2})^{3}}+\frac{8k^{\alpha}k^{\beta}a_{\alpha\beta}}{(k^{2})^{4}}\Big]_{k=k_{1}+k_{2}+k_{3}},
\end{equation} 
which becomes dominant over curvature independent term when $k_{1}+k_{2}+k_{3}\rightarrow0$. On the other hand, for the minimal coupling (when $\xi=0$), leading order correction comes from $\frac{1}{6}\frac{\mathcal{R}}{((k_{1}+k_{2}+k_{3})^{2})^{2}}$ for massless scalar field theory, however, depending on the limit $k_{1}+k_{2}+k_{3}\rightarrow0$, higher order terms become important.

\subsection{$2\rightarrow2$ scattering in $\phi^{3}$ theory}
In this case, we obtain the following expression
\begin{equation}
\begin{split}
\mathcal{S}(k_{1},k_{2}\rightarrow k_{3},k_{4}) & =\int d^{4}z_{1}\ldots d^{4}z_{4}f_{k_{1}}(z_{1})f_{k_{2}}(z_{2})f_{k_{3}}^{*}(z_{3})f_{k_{4}}^{*}(z_{4})V(z_{1},\ldots,z_{4})\\
V(z_{1},\ldots,z_{4}) & =\frac{\delta}{\delta\varphi(z_{1})}\ldots\frac{\delta}{\delta\varphi(z_{4})}\mathcal{F}[\varphi]|_{\varphi=0}, \ \mathcal{F}[\varphi]=e^{\frac{i}{2}\int_{x,y}\delta_{x}\mathcal{G}(x,y)\delta_{y}}e^{-i\frac{\lambda}{3!}\int_{z}\varphi^{3}(z)\sqrt{-g(z)}}.
\end{split}
\end{equation}
If we consider $\varphi^{4}$ term from $\mathcal{F}[\varphi]$, then in the leading order, we obtain the following contribution
\begin{equation}
\begin{split}
\frac{1}{2}\int_{z_{1},z_{2}}\delta_{z_{1}}\mathcal{G}(z_{1},z_{2})\delta_{z_{2}} & \Big[-\frac{\lambda^{2}}{2(3!)^{2}}\left(\int_{x}\varphi^{3}(x)(1-\frac{1}{6}R_{\alpha\beta}x^{\alpha}x^{\beta})\right)^{2}\Big]\\
=-\frac{\lambda^{2}}{16}\int_{z_{1},z_{2},x,y}\bar{\mathcal{G}}(z_{1},z_{2}) & \varphi^{2}(x)\delta^{(4)}(x-z_{1})\delta^{(4)}(y-z_{2})\varphi^{2}(y)\left(1-\frac{1}{12}R_{\alpha\beta}x^{\alpha}x^{\beta}\right)\left(1-\frac{1}{12}R_{\alpha\beta}y^{\alpha}y^{\beta}\right)\\
-\frac{\lambda^{2}}{24}\int_{z_{1},z_{2},x,y}\bar{\mathcal{G}}(z_{1},z_{2}) & \delta^{(4)}(x-z_{1})\delta^{(4)}(x-z_{2})\varphi(x)\varphi^{3}(y)\left(1-\frac{1}{12}R_{\alpha\beta}x^{\alpha}x^{\beta}\right)\left(1-\frac{1}{12}R_{\alpha\beta}y^{\alpha}y^{\beta}\right)\\
-\frac{\lambda^{2}}{24}\int_{z_{1},z_{2},x,y}\bar{\mathcal{G}}(z_{1},z_{2}) & \delta^{(4)}(y-z_{1})\delta^{(4)}(y-z_{2})\varphi^{3}(x)\varphi(y)\left(1-\frac{1}{12}R_{\alpha\beta}x^{\alpha}x^{\beta}\right)\left(1-\frac{1}{12}R
_{\alpha\beta}y^{\alpha}y^{\beta}\right).
\end{split}
\end{equation}
As before we look at the connected piece only which can be described from the Fig.2.
\begin{figure}
\begin{center}
\includegraphics[height=3.7cm,width=4cm]{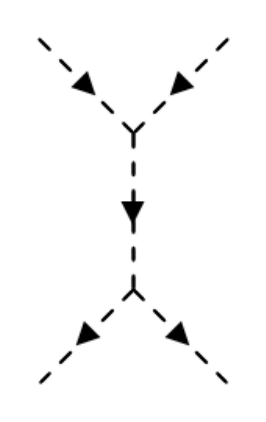}
\end{center}
\caption{Leading order contributing diagram in $2\rightarrow2$ scattering in $\phi^{3}$ theory}
\end{figure}
Here, from the various combinations, we choose a particular one that contributes as
\begin{equation}
\begin{split}
\mathcal{S}^{(1)}(k_{1},k_{2}\rightarrow k_{3},k_{4}) & =-\frac{\lambda^{2}}{4}\int d^{4}x \ d^{4}y \ 
d^{4}k\left(1+\frac{1}{12}(\partial_{k_{1}}^{\alpha}\partial_{k_{1}}^{\beta}+\partial_{k_{3}}^{\alpha}\partial_{k_{3}}^{\beta})+\frac{1}{144}R_{\alpha\beta}R_{\mu\nu}\partial_{k_{1}}^{\alpha}\partial_{k_{1}}^{\beta}\partial_{k_{3}}^{\mu}\partial_{k_{3}}^{\nu}\right)\\
 & \times e^{i(k_{1}+k_{2}-k).x-i(k_{3}+k_{4}-k).y}\frac{\bar{\mathcal{G}}(k)}{\prod_{i=1}^{4}\sqrt{2\omega_{k_{i}}V}}\\
 & =-\frac{\lambda^{2}}{4}\left(1+\frac{1}{12}(\partial_{k_{1}}^{\alpha}\partial_{k_{1}}^{\beta}+\partial_{k_{3}}^{\alpha}\partial_{k_{3}}^{\beta})+\frac{1}{144}R_{\alpha\beta}R_{\mu\nu}\partial_{k_{1}}^{\alpha}\partial_{k_{1}}^{\beta}\partial_{k_{3}}^{\mu}\partial_{k_{3}}^{\nu}\right)\Bigg[\frac{\bar{\mathcal{G}}(k_{1}+k_{2})}{\prod_{i=1}^{4}\sqrt{2\omega_{k_{i}}V}}\\
 & \times\delta^{(4)}(k_{1}+k_{2}-k_{3}-k_{4})\Bigg].
\end{split}
\end{equation}
Like the previous example, here, we have conservation and non-conservation terms, and both contain curvature-dependent corrections. In the conservation case, in the $k_{1}+k_{2}\rightarrow 0$ limit, curvature correction dominates over curvature-independent correction in the massless theory. The IR divergences can be cured by first putting an IR regulator and then adding soft-modes in the theory.

\subsection{Probing curved spacetime features through Nucleon-Nucleon scattering}
Nucleon-Nucleon scattering is a Yukawa process where nucleons exchange mesons. 
\begin{figure}
\begin{center}
\includegraphics[height=5cm,width=7cm]{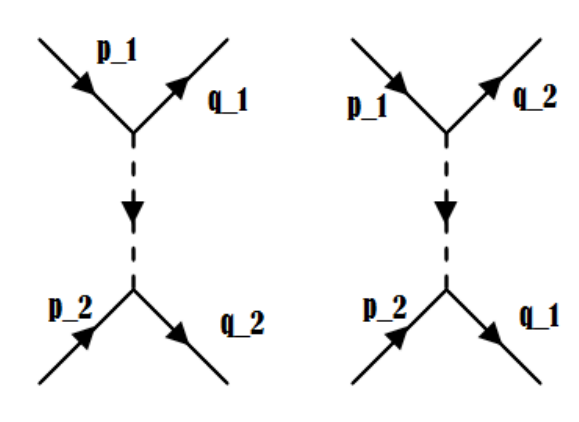}
\end{center}
\caption{Nucleon-Nucleon scattering process diagrams in leading order}
\end{figure}
In the center of mass frame, the incoming energy of the two particles is identical and equal to half the center of mass-energy, while the 3-momenta of the incoming particles are equal and opposite, $p_{1}=-p_{2}$. As energy is conserved, the energy of the out-going particles is also half the center of mass-energy. This means that the three-momenta of the outgoing particles are also equal and opposite $q_{1}=-q_{2}$. Since all the energies are equal and all the masses are equal, it follows that the magnitudes of all the momenta are equal $|\vec{p}_{1}|=|\vec{p}_{2}|=|\vec{q}_{1}|=|\vec{q}_{2}|=|\vec{p}|=p$. The scattering cross-section of this process is given by
\begin{equation}
\frac{d\sigma}{d\Omega}\propto\frac{\lambda^{4}}{\sqrt{p^{2}+M^{2}}}\Big[\frac{1}{2|\vec{p}|^{2}(1+\cos\theta)+m^{2}}+\frac{1}{2|\vec{p}|^{2}(1-\cos\theta)+m^{2}}\Big]^{2},
\end{equation}
where $M$ is the mass of the nucleons and $\lambda$, coupling constant, controls the amount of scattering, i.e., it is a measure of the strength of the interactions. The scattering has a non-trivial dependence on $\theta$ (scattering angle). The theory has a specific prediction for its form which can be used to fit the data to reveal the best value for $m$, the mass of the meson-particles which is not directly measurable. In the limit of small incoming momenta $|\vec{p}|$ compared to the mass of meson, the above formula becomes
\begin{equation}
\frac{d\sigma}{d\Omega}\propto\frac{\lambda^{4}}{\sqrt{p^{2}+M^{2}}m^{4}}.
\end{equation}
The same scattering process in condensed matter systems, where electrons are interacting through the same interaction but exchanging massless phonons, is
\begin{equation}\label{eqn.22}
\frac{d\sigma}{d\Omega}\propto\frac{\lambda^{4}}{\sqrt{p^{2}+M^{2}}}\Big[\frac{1}{2|\vec{p}|^{2}(1+\cos\theta)}+\frac{1}{2|\vec{p}|^{2}(1-\cos\theta)}\Big]^{2}=\frac{\lambda^{4}}{|\vec{p}|^{4}\sqrt{p^{2}+M^{2}}}\frac{1}{sin^{4}\theta},
\end{equation} 
which shows IR-singularity structure of scattering cross-section and also singularity structure in $\theta\rightarrow0,\pi$ limit. The scattering cross-section, in eqn. (\ref{eqn.22}) for 4-momentum conservation piece in curved spacetime in minimal coupling theory becomes (after taking into account the first order curvature dependent term)
\begin{equation}
\begin{split}
\frac{d\sigma}{d\Omega}\propto & \frac{\lambda^{4}}{\sqrt{p^{2}+M^{2}}}\Big[\frac{1}{2|\vec{p}|^{2}(1+\cos\theta)+m^{2}}+\frac{\mathcal{R}}{6}\frac{1}{(2|\vec{p}|^{2}(1+\cos\theta)+m^{2})^{2}}\\
 & +\frac{1}{2|\vec{p}|^{2}(1-\cos\theta)+m^{2}}+\frac{\mathcal{R}}{6}\frac{1}{(2|\vec{p}|^{2}(1-\cos\theta)+m^{2})^{2}}\Big]^{2}.
\end{split}
\end{equation}
In $m\gg|\vec{p}|$ limit, the above equation becomes
\begin{equation}\label{eqn.23}
\frac{d\sigma}{d\Omega}\propto\frac{\lambda^{4}}{\sqrt{p^{2}+M^{2}}}\left(\frac{2}{m^{2}}+\frac{\mathcal{R}}{3m^{4}}\right)^{2}.
\end{equation}
If $\mathcal{R}(x')$ or $\mathcal{R}(z=0)$ is much greater compared to the mass such that the 2nd term in eqn. (\ref{eqn.23}) dominates, then 
\begin{equation}
\frac{d\sigma}{d\Omega}\propto\frac{\lambda^{4}}{\sqrt{p^{2}+M^{2}}}\frac{\mathcal{R}^{2}}{9m^{8}},
\end{equation}
which carries information about the curvature through dependence on Ricci scalar. On the other hand, in the $m\rightarrow0$ limit, we obtain the following expression
\begin{equation}
\frac{d\sigma}{d\Omega}\propto\frac{\lambda^{4}}{\sqrt{p^{2}+M^{2}}}\left(\frac{1}{|\vec{p}|^{2}\sin^{2}\theta}+\frac{\mathcal{R}}{12|\vec{p}|^{4}}\frac{1+\cos^{2}\theta}{sin^{4}\theta}\right)^{2},
\end{equation}
which shows that in $|\vec{p}|\rightarrow0$ limit (IR structure of scattering cross-section without the other higher order curvature terms), the second term  dominates over the first term unless $\mathcal{R}$ is sufficiently small, thereby leading to 
\begin{equation}\label{eqn.24}
\frac{d\sigma}{d\Omega}\propto\frac{\lambda^{4}}{\sqrt{p^{2}+M^{2}}}\frac{\mathcal{R}^{2}}{144|\vec{p}|^{8}}\left(\frac{1+\cos^{2}\theta}{\sin^{4}\theta}\right)^{2}.
\end{equation}
In the $\theta\rightarrow0$ limit, the singularity structure becomes
\begin{equation}\label{eqn.25}
\frac{d\sigma}{d\Omega}\propto\frac{\lambda^{4}}{\sqrt{p^{2}+M^{2}}}\frac{\mathcal{R}^{2}}{36|\vec{p}|^{8}\sin^{8}\theta}.
\end{equation}
Both the limiting cases, eqns. (\ref{eqn.24}) and (\ref{eqn.25}), carry information about Ricci scalar of spacetime at $x'$ where scattering events take place through local interaction.

In the conformal coupling limit $\xi=\frac{1}{6}$, it can be checked easily that in the two limiting cases mentioned above, the scattering cross-section depends on $\frac{1}{|\vec{p}|^{12}\sin^{12}\theta}$ and curvature dependence signature carried by the terms would be  $\mathcal{O}((a_{\alpha}^{ \ \alpha})^{2},a_{\alpha\beta}a_{\mu\nu})$. These observations are basically spacetime geometrical features that one can probe through the local formulation of S-matrix in a Nucleon-Nucleon scattering process.

\subsection{Nucleon interactions in curved spacetime}
It is quite well known that relativistic theories of nucleons interacting with mesons are very useful for explaining a wide range of nuclear phenomena, most notably the single-particle energy levels, nuclear charge densities and elastic proton-nucleus scattering observables at intermediate energies \cite{walecka1986relativistic, wallace1987relativistic, reinhard1989relativistic}.

We consider a scalar theory description of nucleons \cite{dick2007scalar, kraus1992renormalization} for the sake of mathematical simplicity. Nucleon interactions involve the strong nuclear force. These interactions are known as hadronic interactions, and the involved particles in these interactions are called hadrons. In the scalar model, the hadronic interaction is affected by the exchange of scalar mesons between nucleons. More fundamentally, hadronic interactions are mediated by gluons, which are exchanged by the quark constituents of the hadrons. Thus, the nucleon-meson model is seen to be an effective field theory that replaces the fundamental quark-gluon interaction with the simpler effective nucleon-meson interaction.

The simplest description of the interaction is written as a product of three fields $\phi_{A}(x), \  \phi_{B}(x)$ and $\phi_{C}(x)$ and coupling constants $g_{AAC}$ and $g_{ABC}$. The interaction Lagrangian density (in Euclidean formalism) is given by
\begin{equation}
\mathcal{L}_{I}=g_{AAC}\phi_{A}(x)\phi_{A}(x)\phi_{c}(x)+g_{ABC}\phi_{A}(x)\phi_{B}(x)\phi_{C}(x), \ S_{I}=\int\sqrt{-g(x)}\mathcal{L}_{I}(\phi_{A,B,C}(x))d^{4}x,
\end{equation}
where A field represents nucleons, B field the $\Delta$ baryons, and C field the mesons. Though the scalar model recognizes only two particle attributes, this model is sufficient to determine scattering amplitudes, from which decay rates and differential and total cross-sections can be derived. This Lagrangian forms the basis for the work in \cite{norbury2010cross}. This model can be treated as an effective field theory that gives scattering cross-sections of different processes in intermediate energy scales to a very good accuracy without worrying about the internal structure of particles. This is the simplest model where two interaction channels are considered at the tree level.

Although the bare model contains only three-point vertices, effectively this model also gives rise to self-interacting four-point vertices. Therefore, to measure the effective strength of such coupling constant in curved spacetime we include extra pieces in the action which is of the following form
\begin{equation}
\frac{1}{4!}\int\sqrt{-g(x)}(\lambda_{A}\phi_{A}^{4}(x)+\lambda_{B}\phi_{B}^{4}(x)+\lambda_{C}\phi_{C}^{4}(x))d^{4}x.
\end{equation}
We show that this effective interaction coupling strength depends on spacetime curvature. Here we look at a particular case of $\phi_{C}^{4}$ vertex coupling strength due to the effect of three-point interactions in curved spacetime.

The diagrams contributing to this are shown in Fig. 4 drawn only for the 4-momentum conservation piece that arises from the first term in the expansion $\sqrt{-g(z)}$ in terms of $z$. In the figure dashed line denotes the propagator of $\phi_{A}$, the solid line denotes the propagator for $\phi_{C}$ and the wavy line represents the propagator for $\phi_{B}$ fields.

\begin{figure}
\begin{center}
\includegraphics[height=4.7cm,width=5.2cm]{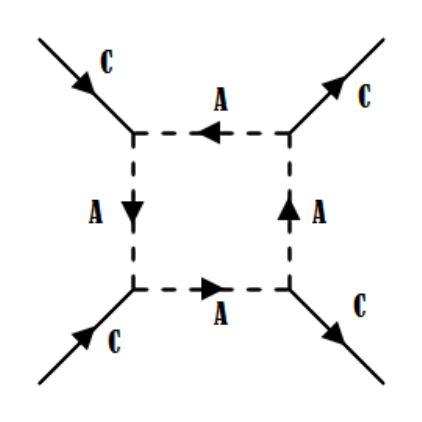}
\includegraphics[height=4.7cm,width=5.2cm]{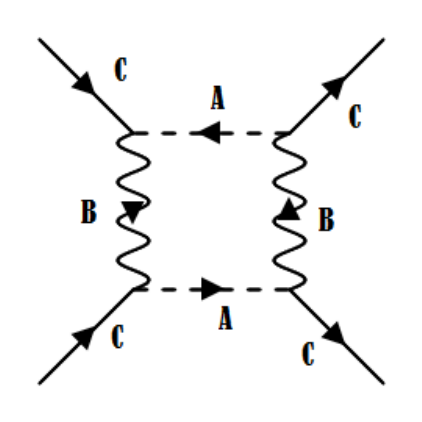}\\
\includegraphics[height=4.7cm,width=5.2cm]{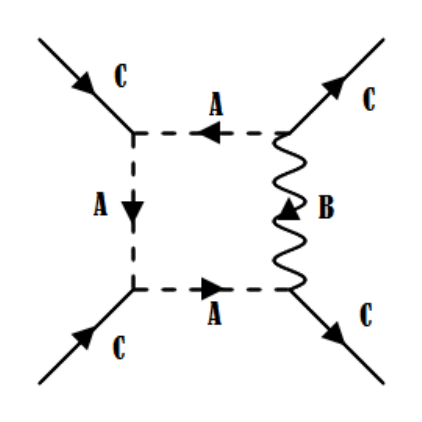}
\caption{i) Four $g_{AAC}$ couplings, ii) four $g_{ABC}$ couplings, iii) two $g_{AAC}$ and two $g_{ABC}$ couplings}
\end{center}
\end{figure}

From Fig. 4.1, we get the following contribution to the effective coupling strength $\lambda_{C}$ (only for the piece where 4-momentum conservation holds)
\begin{equation}
\begin{split}
g_{AAC}^{4} & \int\frac{d^{4}k}{(2\pi)^{4}}\Big[\frac{1}{(k^{2}+M_{A}^{2})^{4}}+\frac{4\mathcal{R}}{6}\frac{1}{(k^{2}+M_{A}^{2})^{5}}+\mathcal{O}(\mathcal{R}^{2},R_{\alpha\beta}R_{\mu\nu})\Big]\\
 & =\frac{g_{AAC}^{4}}{(4\pi)^{2}}\Big[\frac{1}{6M_{A}^{4}}+\frac{\mathcal{R}}{18M_{A}^{6}}+\mathcal{O}\left(\frac{\mathcal{R}^{2}}{M_{A}^{4}},\frac{R_{\alpha\beta}R_{\mu\nu}}{M_{A}^{4}}\right)\Big].
\end{split}
\end{equation}
From Fig. 4.2, we get the following contribution (for the piece where 4-momentum conservation holds)
\begin{align}
g_{ABC}^{4}\int\frac{d^{4}k}{(2\pi)^{4}} & \Bigg[\frac{1}{(k^{2}+M_{A}^{2})^{2}}\frac{1}{(k^{2}+M_{B}^{2})^{2}}+\frac{2\mathcal{R}}{6}\Big[\frac{1}{(k^{2}+M_{A}^{2})^{3}}\frac{1}{(k^{2}+M_{B}^{2})^{2}}+\frac{1}{(k^{2}+M_{B}^{2})^{3}}\frac{1}{(k^{2}+M_{A}^{2})^{2}}\Big] \nonumber\\
 & +\mathcal{O}(\mathcal{R}^{2},R_{\alpha\beta}R_{\mu\nu})\Bigg]\\
=\frac{g_{ABC}^{4}}{(4\pi)^{2}}\Big[\frac{1}{M_{A}^{4}} & \left(\frac{(\zeta+1)\ln\zeta}{(\zeta-1)^{3}}-\frac{2}{(\zeta-1)^{2}}\right)+\frac{\mathcal{R}}{6M_{A}^{6}}\left(-\frac{\ln\zeta}{(\zeta-1)^{3}}+\frac{\zeta+1}{2(\zeta-1)^{2}\zeta}\right)+\mathcal{O}(\mathcal{R}^{2},R_{\alpha\beta}R_{\mu\nu})\Big] \nonumber,
\end{align}
where $\zeta=\frac{M_{B}^{2}}{M_{A}^{2}}$ ratio of renormalized masses of B and A particles. This contribution will add to the previous one. Considering the Fig. 4.3, we get the following contribution
\begin{align}
g_{AAC}^{2}g_{ABC}^{2}\int\frac{d^{4}k}{(2\pi)^{4}} & \Bigg[\frac{1}{(k^{2}+M_{A}^{2})^{3}}\frac{1}{(k^{2}+M_{B}^{2})}+\frac{\mathcal{R}}{6}\Big[\frac{1}{(k^{2}+M_{A}^{2})^{3}}\frac{1}{(k^{2}+M_{A}^{B})^{2}}+\frac{3}{(k^{2}+M_{B}^{2})}\frac{1}{(k^{2}+M_{A}^{A})^{4}}\Big] \nonumber\\
 & +\mathcal{O}(\mathcal{R}^{2},R_{\alpha\beta}R_{\mu\nu})\Bigg]\\
=\frac{g_{AAC}^{2}g_{ABC}^{2}}{(4\pi)^{2}}\Big[\frac{1}{M_{A}^{4}} & \left(\frac{1}{(\zeta-1)^{2}}-\frac{2\zeta\ln\zeta}{(\zeta-1)^{3}}+\frac{\zeta}{(\zeta-1)^{2}}\right)+\frac{\mathcal{R}}{12M_{A}^{6}}\left(\frac{\ln\zeta}{(\zeta-1)^{3}}-\frac{2}{(\zeta-1)^{2}}+\frac{\zeta+1}{2(\zeta-1)^{2}}\right) \nonumber\\
 & +\mathcal{O}(\mathcal{R}^{2},R_{\alpha\beta}R_{\mu\nu})\Big] \nonumber.
\end{align}
Therefore, up to one-loop correction, the effective coupling strength of $\phi_{c}^{4}$ interaction is 
\begin{align}
\frac{\lambda_{C}}{4!} & +\alpha\frac{g_{AAC}^{4}}{(4\pi)^{2}}\Bigg[\frac{1}{6M_{A}^{4}}+\frac{\mathcal{R}}{18M_{A}^{6}}+\mathcal{O}\left(\frac{\mathcal{R}^{2}}{M_{A}^{4}},\frac{R_{\alpha\beta}R_{\mu\nu}}{M_{A}^{4}}\right)\Bigg]+\beta\frac{g_{ABC}^{4}}{(4\pi)^{2}}\Bigg[\frac{1}{M_{A}^{4}}\left(\frac{(\zeta+1)\ln\zeta}{(\zeta-1)^{3}}-\frac{2}{(\zeta-1)^{2}}\right) \nonumber\\
 & +\frac{\mathcal{R}}{6M_{A}^{6}}\left(-\frac{\ln\zeta}{(\zeta-1)^{3}}+\frac{\zeta+1}{2(\zeta-1)^{2}\zeta}\right)+\mathcal{O}\left(\frac{\mathcal{R}^{2}}{M_{A}^{4}},\frac{R_{\alpha\beta}R_{\mu\nu}}{M_{A}^{4}}\right)\Bigg]+\gamma\frac{g_{AAC}^{2}g_{ABC}^{2}}{(4\pi)^{2}}\Bigg[\frac{1}{M_{A}^{4}}\Big[\frac{1}{(\zeta-1)^{2}}-\frac{2\zeta\ln\zeta}{(\zeta-1)^{3}} \nonumber\\
 & +\frac{\zeta}{(\zeta-1)^{2}}\Big]+\frac{\mathcal{R}}{12M_{A}^{6}}\left(\frac{\ln\zeta}{(\zeta-1)^{3}}-\frac{2}{(\zeta-1)^{2}}+\frac{\zeta+1}{2(\zeta-1)^{2}}\right)+\mathcal{O}\left(\frac{\mathcal{R}^{2}}{M_{A}^{4}},\frac{R_{\alpha\beta}R_{\mu\nu}}{M_{A}^{4}}\right)\Bigg],
\end{align}
where $\alpha,\beta,\gamma$ are some numerical factors which arise from combinatorics of the graphs. Although we have computed the leading order curvature dependent correction through this procedure, one can in principle compute all higher-order curvature dependent terms.

This calculation shows how the curvature of spacetime affects the coupling strength of particles in a non-trivial curved spacetime. This shows that the running coupling of any theory living in curved spacetime depends on the geometrical properties of the spacetime itself. Hence, it also controls the renormalization group flow equations, and further, the flow equations vary from point to point on the spacetime manifold.

Through our local S-matrix description, we are able to probe the geometrical features of curved spacetime. Next, we discuss the compatibility of local construction of S-matrix with spacetime symmetries or Killing vector fields of curved spacetime and make some comments related to Unitarity. 

\section{Compatibility of spacetime symmetries with local S-matrix}
In this section, the deformed translational symmetries are derived by solving the Killing equations in a series expansion within the RNC patch. Using these deformed symmetries, it is shown explicitly that the S-matrix elements contain terms violating 4-momentum conservation. This is consistent with our findings in the previous section. Therefore, our result in this section shows that the S-matrix derived using RNC and functional integral formulation is consistent with the deformed symmetries of the RNC patch. 

\subsection{Unitarity, 4-momentum conservation condition of S-matrix in flat spacetime}
Denoting all the incoming-states is denoted by $\ket{\{p_{i}\}}^{(\text{in})}$ and outgoing-states by $\ket{\{q_{j}\}}^{(\text{out})}$, the conservation of probability implies
\begin{equation}
\sum_{\{p_{i}\},\{q_{j}\}}| \ ^{(\text{out})}\bra{\{q_{j}\}}\hat{\mathcal{S}}\ket{\{p_{i}\}}^{(\text{in})}|^{2}=1,
\end{equation}
where $\hat{\mathcal{S}}$ is the S-matrix operator. The above equation can be expressed in the following fashion using the completeness relations of the incoming-states and outgoing-states
\begin{equation}
\begin{split}
1=\sum_{\{p_{i}\}\text{or}\{q_{j}\}}| \ ^{(\text{out})}\bra{\{q_{j}\}}\hat{\mathcal{S}}\ket{\{p_{i}\}}^{(\text{in})}|^{2} & =\sum_{\{p_{i}\}} \ ^{(\text{out})}\bra{\{q_{j}\}}\hat{\mathcal{S}}\ket{\{p_{i}\}}^{(\text{in})} \ ^{(\text{in})}\bra{\{p_{i}\}}\hat{\mathcal{S}}^{\dagger}\ket{\{q_{j}\}}^{(\text{out})}\\
 & =\sum_{\{q_{j}\}} \ ^{(\text{in})}\bra{\{p_{i}\}}\hat{\mathcal{S}}^{\dagger}\ket{\{q_{j}\}}^{(\text{out})} \ ^{(\text{out})}\bra{\{q_{j}\}}\hat{\mathcal{S}}\ket{\{p_{i}\}}^{(\text{in})},\\
\implies\hat{\mathcal{S}}^{\dagger}\hat{\mathcal{S}} & =\hat{\mathcal{S}}\hat{\mathcal{S}}^{\dagger}=\hat{\mathbb{I}}. 
\end{split}
\end{equation}
Thus, the unitarity of the S-matrix follows from the conservation of probability. This derivation is also valid for our local S-matrix description in curved spacetime (see \cite{friedman1992unitarity}). 
 
In Minkowski spacetime, we have four translational symmetries corresponding to the four generators denoted by $\hat{P}^{\mu}$ and defined as
\begin{equation}
\hat{P}^{\mu}=\int_{\Sigma} d^{3}x\hat{T}^{0\mu}(x),
\end{equation} 
where $\Sigma$ is a space-like hypersurface, for example, $t=\text{const.}$ hypersurfaces. Since S-matrix in Minkowski spacetime must be compatible with these spacetime symmetries, this implies that
\begin{equation}
[\hat{\mathcal{S}},\hat{P}^{\mu}]=0, \ \forall\mu\in\{0,1,2,3\}.
\end{equation}
This puts a constraint on the S-matrix elements as follows
\begin{equation}
\begin{split}
0 & = \ ^{(\text{out})}\bra{\{q_{j}\}}[\hat{\mathcal{S}},\hat{P}^{\mu}]\ket{\{p_{i}\}}^{(\text{in})}= \ ^{(\text{out})}\bra{\{q_{j}\}}(\hat{\mathcal{S}}\hat{P}^{\mu}-\hat{P}^{\mu}\hat{\mathcal{S}})\ket{\{p_{i}\}}^{(\text{in})}\\
 & = \ ^{(\text{out})}\bra{\{q_{j}\}}\hat{\mathcal{S}}\ket{\{p_{i}\}}^{(\text{in})}\times(\sum_{i}p_{i}^{\mu}-\sum_{j}q_{j}^{\mu}),\\
\implies &  \ ^{(\text{out})}\bra{\{q_{j}\}}\hat{\mathcal{S}}\ket{\{p_{i}\}}^{(\text{in})}=\delta^{(4)}(\sum_{j}q_{j}-\sum_{i}p_{i})\mathcal{A}(\{p_{i}\}\rightarrow\{q_{j}\}),
\end{split}
\end{equation}
which is nothing but 4-momentum conservation during scattering event of particles. Here $\mathcal{A}(\{p_{i}\}\rightarrow\{q_{j}\})$ is the scattering amplitude for incoming states $\ket{\{p_{i}\}}^{(\text{in})}$ to outgoing states $\ket{\{q_{j}\}}^{(\text{out})}$.

\subsection{Symmetries in curved spacetime and compatibility with S-matrix}\label{symmetry subsection}
Symmetries in curved spacetime are described by Killing vector fields denoted by $\xi^{\mu}$ which satisfy the following equation
\begin{equation}\label{eqn.26}
\nabla_{\mu}\xi_{\nu}+\nabla_{\nu}\xi_{\mu}=0,
\end{equation}
known as the Killing equation \cite{Carroll:1602292, misner2017gravitation}. In a generic curved spacetime, the existence of Killing vector fields or solutions of Killing equations is not expected. However, since we only look for symmetries inside the RNC patch, we only need the vector fields which are solutions of Killing equations inside the RNC patch. The Taylor-series expansion of the metric tensor, Christoffel symbol, Riemann curvature tensors makes it possible to get analytical solutions in the form of a series which is shown below.

Given the stress-energy tensor of matter denoted by $T^{\mu\nu}$ which satisfies covariant conservation law $\nabla_{\mu}T^{\mu\nu}=0$, the generators of spacetime symmetries can be obtained through the zeroth component of the following 4-current
\begin{equation}
j^{\mu}=T^{\mu\nu}\xi_{\nu}.
\end{equation}
This satisfies covariant conservation
\begin{equation}
\nabla_{\mu}j^{\mu}=\nabla_{\mu}T^{\mu\nu}\xi_{\nu}+T^{\mu\nu}\nabla_{\mu}\xi_{\nu}=0,
\end{equation}
where we have used symmetric property of stress-energy tensor in two indices which follows from its definition and covariant conservation of stress-energy tensor.
Therefore, the conserved charge which is associated with a spacetime symmetry and plays the role of the generator of spacetime symmetry is
\begin{equation}
\hat{\mathcal{Q}}=\int_{\Sigma} d^{3}x\sqrt{-g}\hat{T}^{0\nu}\xi_{\nu},
\end{equation}
for each Killing vector field and where $\Sigma$ is constant time, space-like hypersurface. One can show that these charges are time-independent and do not depend on the space-like hypersurface.

The metric tensor inside the chosen patch or chart in Riemann-normal coordinates differs from a Minkowski-like metric; it has pieces that deform the spacetime translational symmetries by adding non-trivial contributions to spacetime translational generators, as shown below. This deformed translation symmetry is the origin of the non-conserved pieces of the S-matrix elements in a general curved spacetime.

Now we look at the deformed spacetime translational symmetries in the patch using RNC through the eqn. (\ref{eqn.26})
\begin{equation}
\begin{split}
\nabla_{\mu}K_{\nu} & +\nabla_{\nu}K_{\mu}=0\implies\partial_{\mu}K_{\nu}+\partial_{\nu}K_{\mu}+2\Gamma_{ \ \mu\nu}^{\lambda}K_{\lambda}=0\\
\implies\partial_{\mu}K_{\nu}(0) & +\partial_{\nu}K_{\mu}(0)+(\partial_{\rho}\partial_{\mu}K_{\nu}+\partial_{\rho}\partial_{\nu}K_{\mu})(0)z^{\rho}+2\partial_{\rho}\Gamma_{ \ \mu\nu}^{\lambda}(0)z^{\rho}K_{\lambda}(0)+\mathcal{O}(z^{\mu}z^{\nu})=0.
\end{split}
\end{equation}
According to the first approximation, $K_{\nu}^{(\mu)}(0)=\delta_{\nu}^{\mu}$ which is spacetime translational symmetry in Minkowski spacetime. Therefore, the above equation is reduced to
\begin{equation}
\partial_{\rho}(\partial_{\mu}K_{\nu}^{(\lambda)}+\partial_{\nu}K_{\mu}^{(\lambda)})(0)=-2\partial_{\rho}\Gamma_{ \ \mu\nu}^{\lambda}(0),
\end{equation}
which implies that
\begin{equation}
K_{\mu}^{(\lambda)}(z)=\delta_{\mu}^{\lambda}-\frac{1}{6}(R_{ \ \mu\rho\nu}^{\lambda}+R_{ \ \nu\rho\mu}^{\lambda})(0))z^{\nu}z^{\rho}+\mathcal{O}(z^{\mu}z^{\nu}z^{\sigma}),
\end{equation}
from which it can be seen that
\begin{equation}
\begin{split}
\hat{\mathcal{Q}}^{(\lambda)} & =\int_{\Sigma} d^{3}z\sqrt{-g(z)}\hat{T}^{0\nu}(z)K_{\nu}^{(\lambda)}(z)=\hat{\mathcal{Q}}_{0}^{(\lambda)}+\hat{\mathcal{Q}}_{1}^{(\lambda)}+\ldots,\\
\hat{\mathcal{Q}}_{0}^{(\lambda)}=\hat{P}^{\lambda}, & \  \hat{\mathcal{Q}}_{1}^{(\lambda)}=-\frac{1}{3}R_{\alpha\beta}\int_{\Sigma} d^{3}z \ \hat{T}^{0\lambda}(z)z^{\alpha}z^{\beta}-\frac{1}{6}R_{ \ \mu\rho\nu}^{\lambda}\int_{\Sigma} d^{3}z \ \hat{T}^{0\nu}(z)z^{\mu}z^{\rho}.
\end{split}
\end{equation}
Similarly, the local S-matrix can be decomposed in the  following manner
\begin{equation}
\hat{\mathcal{S}}=\hat{\mathcal{S}}_{0}+\hat{\mathcal{S}}_{1}+\ldots,
\end{equation}
where $\hat{\mathcal{S}}_{0}$ is the piece where 4-momentum conservation holds. Therefore, it satisfies
\begin{equation}
[\hat{\mathcal{S}}_{0},\hat{\mathcal{Q}}_{0}]=0.
\end{equation}
However, we claim the following condition holds more generally
\begin{equation}
[\hat{\mathcal{S}},\hat{\mathcal{Q}}^{(\lambda)}]=0=[\hat{\mathcal{S}}_{0},\hat{\mathcal{Q}}_{1}^{(\lambda)}]+[\hat{\mathcal{S}}_{1},\hat{\mathcal{Q}}_{0}^{(\lambda)}]+\ldots, \ \forall\lambda,
\end{equation}
where we write down the terms in order of derivatives of metric. This means
\begin{equation}
[\hat{\mathcal{S}}_{0},\hat{\mathcal{Q}}_{1}^{(\lambda)}]+[\hat{\mathcal{S}}_{1},\hat{\mathcal{Q}}_{0}^{(\lambda)}]=0, \ \forall\lambda,
\end{equation}
which puts the following constraint on the matrix elements of the $\hat{\mathcal{S}}_{1}$ operator
\begin{equation}
^{(\text{out})}\bra{\{q_{j}\}}[\hat{\mathcal{S}}_{0},\hat{\mathcal{Q}}_{1}^{(\lambda)}]\ket{\{p_{i}\}}^{(\text{in})}=- \ ^{(\text{out})}\bra{\{q_{j}\}}[\hat{\mathcal{S}}_{1},\hat{\mathcal{Q}}_{0}^{(\lambda)}]\ket{\{p_{i}\}}^{(\text{in})}= \ ^{(\text{out})}\bra{\{q_{j}\}}\hat{\mathcal{S}}_{1}\ket{\{p_{i}\}}^{(\text{in})}\left(\sum_{j}q_{j}^{\lambda}-\sum_{i}p_{i}^{\lambda}\right).
\end{equation}
Hence, it follows that
\begin{equation}
\begin{split}
\Bigg[\sum_{\{\bar{p}_{k}\}} \ ^{(\text{out})}\bra{\{q_{j}\}}\hat{\mathcal{S}}_{0}\ket{\{\bar{p}_{k}\}}^{(\text{in})} & \ ^{(\text{in})}\bra{\{\bar{p}_{k}\}}\hat{\mathcal{Q}}_{1}^{(\lambda)}\ket{\{p_{i}\}}^{(\text{in})}-\sum_{\{\bar{q}_{l}\}} \ ^{(\text{out})}\bra{\{q_{j}\}}\hat{\mathcal{Q}}_{1}^{(\lambda)}\ket{\{\bar{q}_{l}\}}^{(\text{out})} \ ^{(\text{out})}\bra{\{\bar{q}_{l}\}}\hat{\mathcal{S}}_{0}\ket{\{p_{i}\}}^{(\text{in})}\Bigg]\\
= \ ^{(\text{out})}\bra{\{q_{j}\}}\hat{\mathcal{S}}_{1}\ket{\{p_{i}\}}^{(\text{in})} & \left(\sum_{j}q_{j}^{\lambda}-\sum_{i}p_{i}^{\lambda}\right),
\end{split}
\end{equation}
which implies
\begin{equation}
\begin{split}
\Bigg[\sum_{\{\bar{p}_{k}\}}\mathcal{A}(\{\bar{p}_{k}\}\rightarrow\{q_{j}\})\delta^{(4)}\left(\sum_{j}q_{j}-\sum_{k}\bar{p}_{k}\right) &  \ ^{(\text{in})}\bra{\{\bar{p}_{k}\}}\hat{\mathcal{Q}}_{1}^{(\lambda)}\ket{\{p_{i}\}}^{(\text{in})}\\
-\sum_{\{\bar{q}_{l}\}}\mathcal{A}(\{p_{i}\}\rightarrow\{\bar{q}_{l}\})\delta^{(4)}\left(\sum_{i}p_{i}-\sum_{l}\bar{q}_{l}\right)  & \ ^{(\text{out})}\bra{\{q_{j}\}}\hat{\mathcal{Q}}_{1}^{(\lambda)}\ket{\{\bar{q}_{l}\}}^{(\text{out})}\Bigg]\\
= \ ^{(\text{out})}\bra{\{q_{j}\}}\hat{\mathcal{S}}_{1}\ket{\{p_{i}\}}^{(\text{in})} & \left(\sum_{j}q_{j}^{\lambda}-\sum_{i}p_{i}^{\lambda}\right).
\end{split}
\end{equation}
In general, we should expect that the \textit{l.h.s} of the above equation is non-zero which implies that $^{(\text{out})}\bra{\{q_{j}\}}\hat{\mathcal{S}}_{1}\ket{\{p_{i}\}}^{(\text{in})}\neq 0$ even when $\sum_{j}q_{j}\neq\sum_{i}p_{i}$, implying violation of 4-momentum conservation in the $\hat{\mathcal{S}}_{1}$ piece of the local S-matrix. 

In a similar fashion, to each order of derivative of the metric, a series of conditions between $\hat{\mathcal{S}}_{i}$ and $\hat{\mathcal{Q}}_{i}^{(\lambda)}$ can be found which put constraints on the matrix elements of $\hat{\mathcal{S}}_{i}$. Hence, it can be seen that the deformed translational symmetries of curved spacetime allow the non-conservation of 4-momentum in the S-matrix.

\section{Examples of spacetime whose features can be probed}
As seen above, the leading order correction of the scattering amplitude and cross-section depends on the Ricci scalar of the underlying geometry. The Ricci scalar is a scalar invariant, hence, it has the same value in all coordinate systems. This implies we cannot make a non-zero Ricci scalar zero just by choosing coordinates in which the manifold looks locally flat. Curved spacetimes may have a zero Ricci scalar, for example, it is zero for any vacuum solution like the Schwarzschild metric, but if 
spacetime has a non-zero Ricci scalar then it is non-zero in all coordinates.

In the standard model of cosmology, the geometry of our Universe is described by the FRW metric, given by
\begin{equation}
ds^{2}=-dt^{2}+a^{2}(t)\left(\frac{dr^{2}}{1-kr^{2}}+r^{2}(d\theta^{2}+\sin^{2}\theta d\varphi^{2})\right),
\end{equation}
where $k$ is a constant representing the spatial curvature and $a(t)$ is the scale factor of the Universe. The cosmological constant $\Lambda$ is considered to be important in cosmology, characterized by the energy contribution coming from the vacuum fluctuations. This explains the acceleration of the present Universe. The scale factor of the Universe can be determined exactly by solving the Friedmann equations, assuming the matter of the Universe is described by an ideal fluid. The correlation functions in a field theory can be computed using the Green's function in a perturbative manner, and it is shown in the present article that the Green's function in curved spacetime can be expressed as a series expansion with curvature-dependent coefficients using the RNC. Therefore, we expect that the nature of the scale factor of the Universe can be determined from the precise measurements of the correlation functions, which could be challenging. Moreover, the LSZ theorem connects the correlation functions with the S-matrix elements. As a result, we expect the precise measurements of scattering of light, neutrinos, and cosmic rays \cite{seckel1998neutrino, yan2004cosmic, cyr2014limits} could in principle determine the evolution of the scale factor of the Universe.

The interior geometry of a spherically symmetric static star is given by the following metric
\begin{equation}
ds^{2}=-e^{\nu(r)}dt^{2}+e^{\lambda(r)}dr^{2}+r^{2}(d\theta^{2}+\sin^{2}\theta d\varphi^{2}),
\end{equation}
which solves the TOV equations \cite{smith2012tolman, carloni2018covariant, carloni2018covariant1}. The Ricci scalar of this geometry depends on $\nu(r), \lambda(r)$, their derivatives, and it also depends explicitly on the radial coordinate $r$. The equation of state of matter inside these stars can be obtained by computing the expectation value of the stress-energy tensor $\langle\hat{T}_{\mu\nu}\rangle$ \textit{w.r.t} the vacuum state. This can be obtained from the correlation function of matter fields. Therefore, as earlier, we expect precise measurements of scattering and the correlation functions could in principle determine the properties of the equation of state. Moreover, from the measurement of scattering amplitude, it is possible to determine the effective potential around the nucleons inside these stars, which is discussed in \cite{nuoffer1998equation, bartnitzky1996microscopic}. This can also be achieved using the X-ray polarimetry which is suggested in \cite{costa2001efficient, weisskopf2009x, krawczynski2011scientific, beilicke2014design}, as it can measure the scattering properties of a target. The equation of state of matter is possible to obtain by finding the energy per nucleons with the above-mentioned effective potential using the variational approach as discussed in \cite{togashi2016equation, togashi2013variational}. This shows that the computation of S-matrix elements \cite{reddy1997neutrino, reddy1998neutrino} and correlation functions in curved spacetime using the approach mentioned in the present article could be useful in probing the properties of the equation of states in compact astrophysical stars.

\section{Conclusion}
A local S-matrix is defined using RNC coordinates in a general curved spacetime. This facilitates the use of familiar QFT techniques in flat spacetime to compute scattering amplitudes, cross-sections in curved spacetime. Using this formalism, some examples of scattering amplitude computations are shown. Results suggest that in the context of scattering between mesons, interactions between nucleons can probe certain features of a general curved spacetime by looking at the IR structure of scattering cross-sections and measuring interaction strength locally on the spacetime manifold. We have also shown how features of cosmological models and the equation of state of matter in the interior of a star can be probed through the measurements of correlation functions and scattering processes. It is also shown that our constructed local S-matrix contains pieces in which 4-momentum conservation is not valid during scattering events, compatible with spacetime symmetries discussed in section (\ref{symmetry subsection}), although Unitarity follows in the same way as it does in flat spacetime. One of the important criteria is that in the zero curvature limit any description of the S-matrix must reproduce familiar results in flat spacetime which is indeed satisfied in our construction. Analyticity of the local S-matrix remains to be checked, to which we hope to return in the near future. With the present construction of local S-matrix, we have attempted to overcome the problems raised in \cite{marolf2013perturbative, ashtekar1975quantum, dohse2015complex, einhorn2003interacting, bousso2005cosmology,hellerman2001string, fischler2001acceleration, spradlin2002vacuum} as also discussed in the section \ref{Introduction}. Moreover, this construction of local S-matrix can be defined covariantly in a generic curved spacetime with some approximations.

Although QFT in curved spacetime is a semi-classical description of quantum gravity, we believe that this construction of local S-matrix might be helpful in understanding S-matrix construction in full quantum gravity. 

\section{Acknowledgement}
SM wants to thank IISER Kolkata for supporting this work through a doctoral fellowship. He also wants to thank IIT-Jodhpur for their kind hospitality during his stay there, where a part of the work and relevant discussions were done.

\bibliographystyle{unsrt}
\bibliography{draft}

\end{document}